\documentclass[lettersize,journal]{IEEEtran}
\usepackage{amsmath,amssymb,amsfonts}
\usepackage{algorithm, algorithmicx, algpseudocode}
\usepackage{array}
\usepackage[export]{adjustbox}
\usepackage[caption=false,font=footnotesize,labelfont=sf,textfont=sf]{subfig}
\usepackage{textcomp}
\usepackage{stfloats}
\usepackage{url}
\usepackage{verbatim}
\usepackage{graphicx}
\usepackage{bm}
\usepackage{setspace}
\usepackage{multirow}
\usepackage{booktabs}
\usepackage{makecell}
\makeatletter
\let\NAT@parse\undefined
\makeatother
\usepackage{hyperref}  
\graphicspath{{Figures/}}
\def\BibTeX{{\rm B\kern-.05em{\sc i\kern-.025em b}\kern-.08em
    T\kern-.1667em\lower.7ex\hbox{E}\kern-.125emX}}
\usepackage{balance}
\begin{document}
\title{Colour alignment for relative colour constancy via non-standard references}
\author{Yunfeng Zhao, Stuart Ferguson, Huiyu Zhou, Chris Elliott, and Karen Rafferty
\thanks{Y. Zhao is with the School of Electronics, Electrical Engineering \& Computer Science, Queen's University Belfast, United Kingdom, and was also with Institute for Global Food Security, School of Biological Science, Queen's University Belfast, United Kingdom. E-mail: yzhao25@qub.ac.uk }
\thanks{S. Ferguson and K. Rafferty are with the School of Electronics, Electrical Engineering \& Computer Science, Queen's University Belfast, United Kingdom. E-mail: r.ferguson@ee.qub.ac.uk;k.rafferty@qub.ac.uk}
\thanks{H. Zhou is with the School of Computing and Mathematical Sciences, University of Leicester, United Kingdom. E-mail: hz143@leicester.ac.uk}
\thanks{C. Elliott is with the Institute for Global Food Security, School of Biological Science, Queen's University Belfast, United Kingdom, and with the School of Food Science and Technology, Faculty of Science and Technology, Thammasat University, Thailand. E-mail: chris.elliott@qub.ac.uk}
\thanks{Manuscript created November 2021; revised July 2022; accepted September 2022.}}

\maketitle

\begin{abstract}
  Relative colour constancy is an essential requirement for many scientific imaging applications. However, most digital cameras differ in their image formations and native sensor output is usually inaccessible, e.g., in smartphone camera applications. This makes it hard to achieve consistent colour assessment across a range of devices, and that undermines the performance of computer vision algorithms. To resolve this issue, we propose a colour alignment model that considers the camera image formation as a black-box and formulates colour alignment as a three-step process: camera response calibration, response linearisation, and colour matching. The proposed model works with non-standard colour references, i.e., colour patches without knowing the true colour values, by utilising a novel balance-of-linear-distances feature. It is equivalent to determining the camera parameters through an unsupervised process. It also works with a minimum number of corresponding colour patches across the images to be colour aligned to deliver the applicable processing. Three challenging image datasets collected by multiple cameras under various illumination and exposure conditions, including one that imitates uncommon scenes such as scientific imaging, were used to evaluate the model. Performance benchmarks demonstrated that our model achieved superior performance compared to other popular and state-of-the-art methods.
\end{abstract}


\section{Introduction}
\IEEEPARstart{T}{he} rapid emergence of a great variety of portable digital cameras including smartphones has opened opportunities for on-site and consumer-oriented colorimetry \cite{Fan2021}, spectroscopy \cite{Bowden2021}, and imaging \cite{Banik2021} applications such as biosensing \cite{Huang2018} and medical imaging \cite{Hunt2021}. 

Despite the well-established adjustment processes embedded in modern digital cameras such as White Balance (WB), there are still colour distortions due to the very variable camera properties and quality. This occurs because of differences in the manufacturing process of optical sensors by different manufacturers and differences in the algorithms applied during the processing of native sensor data into the final images. As a result, it has bought the challenge of achieving accurate colour interpretation across a range of devices and imaging environments.

The concept of \emph{relative colour constancy (RCC)} was proposed in a previous work with the aim of quantifying the ability to align colours of same objects between images independent of illumination and camera \cite{Zhao2018}. It was developed from the \emph{colour constancy (CC)} concept where only the illumination was considered \cite{Ebner2007}. To achieve RCC, it is important to understand the image formation in a commercial digital camera.

\begin{figure*}[!t]
    \centering
    \includegraphics[width=0.8\textwidth]{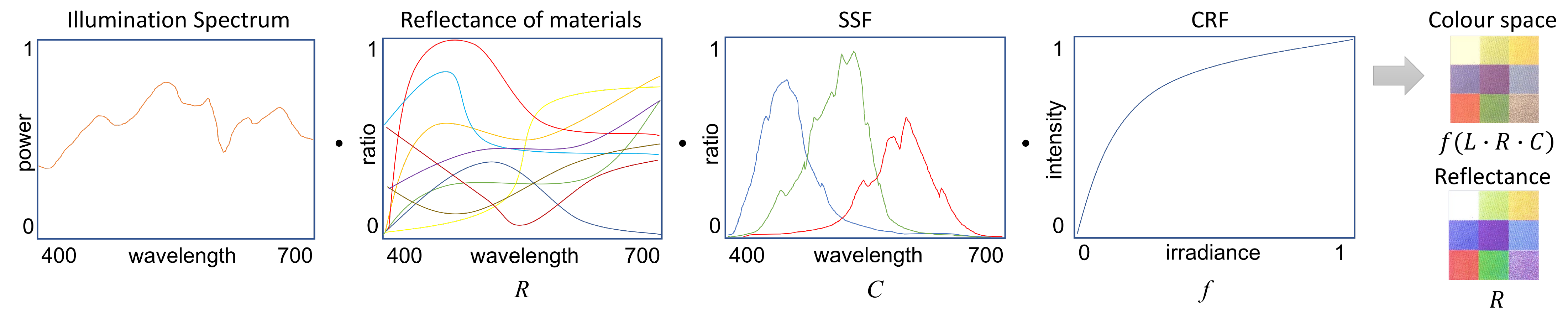}
    \caption{An illustration of the RGB image formation model as a product of illumination, material reflectance, SSF, and CRF. The final image is in a perceptual colour space which is compared with the albedo (reflectance of materials) in the spectral domain.}
    \label{fig:image-formation}
\end{figure*}

Essentially, image formation can be generalized into two phases: optical and signal processing phases. In the optical phase, light emitted from the source of illumination is partially absorbed by materials presented in the imaged scene \cite{Ebner2007}. Light reflected from the scene is focussed through a lens system and Bayer filtered into three colour channels before reaching the image sensor on most commercial RGB cameras. In the signal processing phase, the RAW sensor output is initially generated as a function of the light irradiation received. It is then processed by image post-processing procedures where often nonlinear transforms are applied to the intensity of each colour channel before producing the final image which is usually in JPEG format \cite{Karaimer2016}. The illumination power distribution $E$ received by the image sensor of a typical digital camera at a specific spatial location $x$ can be modelled as the product of the spectral power distribution of the illumination $L$, and albedo of the scene Lambertian surfaces $R$ at the corresponding location, and sensor spectral sensitivity function (SSF) $C$ \cite{Jiang2013, Barnard2002}:

\begin{equation}\label{equ:image-formation}
    {E}\left( x \right) = \int {L\left( {\lambda ,x} \right)R\left( {\lambda ,x} \right){C}\left( \lambda  \right)d\lambda }  + \varepsilon_1
\end{equation}

Next, the illumination power distribution $E$ processed by the image sensor is transformed to the RAW pixel intensities through a sensor response function (SRF) $s$ \cite{Egan2008}. The final JPEG or PNG image $D$ is colour transformed by applying numerous image post-processing procedures such as WB, demosaic, gamma correction, gamut mapping, colour rendering, and data compression as denoted by $o$ \cite{Karaimer2016}. The combined effect of $s$ and $o$ is known as the camera response function (CRF) and is denoted by $f$:

\begin{gather}\label{equ:camera-response}
    RAW = s\left( E  \right) + \varepsilon_2 \\
    D = o\left( {RAW} \right) + \varepsilon_3 \\
    D = f\left( E  \right) + \varepsilon_4 = o\left(s\left( E  \right)+ \varepsilon_2 \right) + \varepsilon_3
\end{gather}

The random noise $\varepsilon$ in every phase of the imaging pipeline cannot be ignored due to the limited camera quality. The image formation pipeline and the produced image are visually illustrated in Fig. \ref{fig:image-formation}.

Modern commercial digital cameras are usually equipped with multiple versions of image post-processing procedures such as scene and portrait mode. In this context, ’camera’ refers to as the combination of camera hardware and fixed image post-processing procedure. The same camera hardware coupled with different image post-processing procedures are considered different cameras. 

There are other user changeable camera properties such as camera lens, exposure, aperture, and image sensor gain and offset that affect the amount of light processed by image sensor. Also, unfortunately, image post-processing algorithms are proprietary information for most commercial digital cameras such as smartphones and are commercially sensitive and unpublished, making step-by-step correction infeasible \cite{Badano2014}. This handicaps their naive use as scientific measuring instruments.

To make scientific measurements from images acquired with digital cameras/smartphones, an image processing procedure that delivers image consistency across a wide range of devices based on only the final JPEG/PNG images is required. Camera response calibration where camera properties are calibrated in advance for later correction is a widely applied strategy for achieving RCC \cite{Zhao2018, Gong2016}. Most camera calibration methods are once-and-for-all methods, i.e., one calibration is done and the calculated properties reused for later corrections. Standard calibrations require professional operations and the use of standard colour references such as standard colour charts and monochromators with true colour or spectral values \cite{ISO-17321}. This way, the colour outcome of the calibrated cameras can be aligned with a standard or the physical nature of the light to achieve a consistent colour performance across calibrated cameras. However, these requirements have clearly become an obstacle for the calibration of consumer-oriented portable cameras as standard colour references are expensive and inaccessible for ordinary consumers. Also, standard colour charts can hardly be fitted in scientific imaging scenes with microscope or light-box. Colour matching where the colour of images are mapped with each other \cite{Gong2016} is another popular strategy for achieving RCC. Existing colour matching methods require a large number of corresponding colour patches (CCPs) across the images in order to achieve an accurate colour transformation from base images to corrected images \cite{Westland2012}. However, it is unusual to find a consistent set of CCP locations across a range of images. A minimisation of the number of CCPs (NoCCPs) required for colour matching is needed before a transformation can be deployed in the field.

This paper presents a colour alignment model that calibrates camera properties and corrects and matches colour of images taken by different cameras and under varied illuminations to achieve RCC. The model presumes inaccessible camera properties, internal image formation, and raw sensor data. And it demonstrates high applicability to be used on portable devices. The work makes the following principal contributions: 
\begin{enumerate}
    \item The proposed colour alignment is modelled as a three-step process: camera response calibration, response linearisation, and colour matching.
    \item The model works with non-standard reference, e.g., colour charts without true colour values, and in an unsupervised manner by utilising a novel balance-of-linear-distances (BoLD) feature.
    \item It is applicable to be used on portable device and operated by ordinary consumers as it can work with a minimum requirement of utilising only four calibration images and two CCPs for colour matching.
    \item Finally, thorough evaluations and benchmarks of the proposed model on three image datasets, i.e., two image datasets of common scenes and an image dataset of uncommon scenes mimicking scientific imaging scenes, have been conducted, and their results indicate that the proposed model achieved the state-of-the-art RCC performance.
\end{enumerate}

\section{Related Works}

\subsection{White Balance}

WB eliminates colour drift effects due to varied illumination colours. It is the most widely used technique for achieving CC. WB usually works in a two-step procedure: estimating the colour of the light source and then compensating the colour drift due to the illumination colours \cite{Ebner2007}.

Illumination colour estimated from a single image can be classified as arising from two analytical approaches: statistical and learning-based methods. Statistical methods estimate the illumination colour mostly by statistical assumptions on a single image. For instances, the Grey-World \cite{Ebner2007} assumes that the world is grey on average and estimates the illumination colour by combining the mean intensity of each colour channel. Similarly, the White-Patch combines the maximum of each colour channel to estimate illumination colour \cite{Ebner2007}. The Double-Opponent \cite{Gao2015} achieves the estimation by mimicking the physiological mechanism in human vision system. Other statistical estimation methods include Shade-of-Grey \cite{Finlayson2004}, Grey-Edge \cite{VandeWeijer2007} and Spatial correlations \cite{Chakrabarti2012}. The major strengths of these statistical methods are their computational efficiency and direct operation that does not need a training or calibration process. The learning-based methods exploit more comprehensive assumptions through a training phase. The Bag-of-Colour-Feature automatically selects the optimal statistical features for illumination colour estimation from training images \cite{Laakom2019}. The Bayesian approach estimates the illumination colour and reflectance of materials from the posterior distribution of colour intensities \cite{Gehler2008}. Other learning-based methods for achieving CC include those using k-nearest neighbor \cite{Afifi2019}, convolutional neural networks (CNN) \cite{Barron, Bianco2015, afifi2019SIIE, yu2020cascading}, generative adversarial networks (GANs) \cite{Das2018}, convolutional autoencoder (CAE) \cite{Laakom2019-2, Afifi2020, Afifi2021}, etc.

In illumination colour correction, WB assumes the camera obeys the von Kries hypothesis, i.e., linear independent intra- and inter-sensor spectral sensitivities among all the colour channels \cite{Cheng2015}. The illumination colour drift in images taken by such cameras can be compensated by multiplying a $3\times3$ diagonal matrix across each pixel on the produced images. The diagonal values of the matrix are the channel-wise ratios between the perfect white and white-point (i.e., the illumination colour) of the image.

\subsection{Camera response calibration}

Referenced and radiometric calibration are the two major categories of methods for camera response calibration. Referenced calibration methods mostly use standard colour references, e.g., standard colour charts and monochromators. The recommended operations to perform the referenced calibration are standardized in ISO 17321-1:2012 \cite{ISO-17321}. Radiometric calibration measures the nonlinearity in colour intensity. The nonlinearity in colour intensity is mainly caused by the nonlinear CRF transformations during the image formation. Its presence reduces the performance of computer vision tasks that require linearity of colour intensity such as image mosaic \cite{Kim2008}, high dynamic range imaging \cite{Reinhard}, and deblurring \cite{Tai2013}. To estimate the CRF, the most accurate way is to image a standard grey chart \cite{Grana2004} or a steady scene with multiple known exposures \cite{Mann2000, Grossberg2003}. There are radiometric calibration methods that measure the colour intensity nonlinearity in unsupervised manners that without using the absolute properties such as true colour values and pre-set camera exposures. Such methods work by analysing histogram \cite{Lin2005}, geometry invariants \cite{Ng2007}, and colour blending \cite{Lin2004} in edge regions, and by using probabilistic intensity similarity \cite{Takamatsu2008} and noise distribution in each image \cite{Matsushita2007}. Deep learning has also been utilised by numeral recent works \cite{Sharma2020, Li2017}.

A major obstacle in camera radiometric calibration is the ambiguity between $E$, $D$, and CRF in the image formation model \cite{Kim2008, Lin2004}. The root of this ambiguity is due to the immeasurability of $E$ that might have been scaled or offset by some value and inaccessible camera properties such as exposure and aperture settings in the image formation. This ambiguity is usually solved by presuming priori regularizations and constraints such as CRF monotonicity and smoothness \cite{Grossberg2003, Rodrigues2015}.

\subsection{Camera response nonlinearity}
\label{subsection:emor}

Nonlinearity in a camera response due to nonlinear transformations by optics and software processing is well known. Camera responses can be decoupled into colour intensity and chromaticity components. Colour intensity is defined as the sum of all the colour channels. Blue-and-red (BR) chromaticity of an RGB colour space is denoted $b$ and $r$:

\begin{gather}\label{equ:chromaticity-b}
    b = \frac{B}{I} = \frac{B}{{R + G + B}} \\
    r = \frac{R}{I} = \frac{R}{{R + G + B}} \\
    g = \frac{G}{I} = \frac{G}{{R + G + B}} = 1 - r - b
\end{gather}

Colour intensity linearity is defined as the one-dimensional space that is linear to the light irradiance in the spectral domain. The most widely adopted camera response nonlinear model is the non-deterministic empirical response model (EMoR) \cite{Grossberg2004}. In the model, a small number of eigenvectors that represent the space of CRF were calculated by applying principal component analysis (PCA) to a database of 201 real-world CRFs obtained from film and digital cameras (DoRF). A CRF represented by the EMoR has the form:

\begin{equation}\label{equ:dorf}
    f=H_0+\bm{\theta}^T H_{1:k}
\end{equation}
where $H_0$ is the mean of all CRFs in the DoRF, $\bm{\theta}$ is the coefficients, and $H_{1:k}$ is the first $k$ eigenvectors with largest eigen values.

Other popular nonlinear CRF models reported in the literature include the polynomial \cite{Mitsunaga1999} and the generalized gamma curve model (GGCM) \cite{Ng2007}.

Chromaticity linearity has not been well studied nor specifically defined. One possible definition of chromaticity linearity is the Luther condition, i.e., the colour spaces that are linearly correlated to human visual sensitivities such as the CIE colour spaces \cite{Karaimer2016, Westland2012}. Since human colour perception is complex and hard to quantify, the main significance of this definition is the accurate colour reproduction for human visual interpretation, and yet it is applicable for chromaticity nonlinear modelling. Another possible definition of the chromaticity linearity is the adaptation of the von Kries hypothesis, i.e., a linear and independent intra- and inter-sensor spectral sensitivity among all the colour channels \cite{Cheng2015}.

\section{Proposed Method}

In this section, a three-step colour alignment framework is proposed. The camera response calibration and colour matching steps in the framework are also explained.

\subsection{Colour alignment framework}

\begin{figure*}[htbp]
    \centering
    \includegraphics[width=0.9\linewidth]{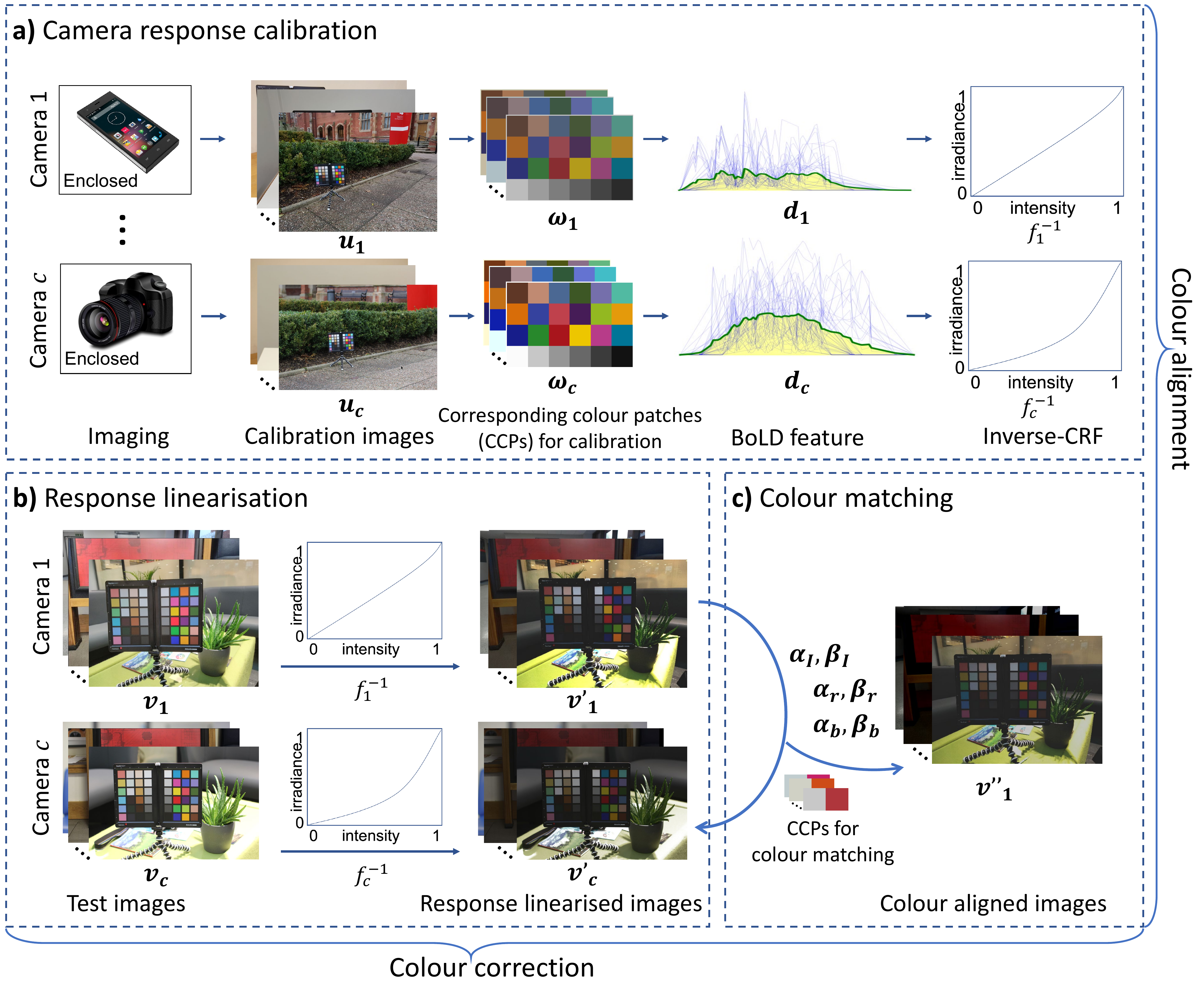}
    \caption{The three-step colour alignment framework. The framework mainly consists of camera response calibration, response linearisation, and colour matching steps. $\bm{u}$ and $\bm{v}$ denote calibration and test images, respectively. $\bm{v}'$ denotes the colour-linearised images. $\bm{v}''$ denotes colour-aligned images. $f^{-1}$ represents the calibrated inverse-CRF. $\bm{\alpha}$ and $\bm{\beta}$ are the transformation coefficients.}
    \label{fig:framework}
\end{figure*}

As illustrated in Fig. \ref{fig:framework}, the proposed colour alignment framework consists of three main steps: camera response calibration, response linearisation, and colour matching. The camera response calibration estimates the inverse of the nonlinearity in colour intensities measured by a camera w.r.t the scene radiance. The test images taken by that camera are colour-linearised by the nonlinear camera response calibrated. In the final step, the intensity and chromaticity of the \emph{colour-linearised images} taken by different cameras are aligned to each other to produce the \emph{colour-aligned images}.

The proposed framework can also be formulated mathematically. Given a list of images, the goal of colour alignment is to determine the optimal inverse-CRF (iCRF) in the camera response calibration step and minimise the colour intensity and chromaticity variation between images in the response linearisation and colour matching steps:

\begin{equation}\label{equ:camera-response-calibration}
    \mathop{\arg\min}_{\bm{\theta}} \mathcal{J}\left(f^{-1}_{\bm{\theta}} \left(\bm{u}\right)\right) \\
\end{equation}
\begin{equation}\label{equ:camera-reponse-correction}
    \bm{v'} = {f_{\bm{\theta}}^{ - 1}}\left( \bm{v} \right) = \left[ {\begin{array}{*{20}{c}}
                    {f_{\bm{\theta}}^{-1}\left(D_1\right)} \\
                    \vdots                   \\
                    {f_{\bm{\theta}}^{-1}\left(D_m\right)}
                \end{array}} \right]
\end{equation}
where $\bm{\theta}$ is the optimal camera-dependent parameters to be calibrated, $\mathcal{J}$ represents the colour deviations between images and is calculated as shown by Eq. \ref{equ:processed-CCP-vector}-\ref{equ:caculate-mean-intensity-distance}, $f_{\bm{\theta}}^{-1}$ is the parametric reconstruction of iCRF for a specific camera, $\bm{u}$ and $\bm{v}$ are the calibration and test image vector, respectively.

For the quantification of colour deviations between images, a two-dimensional CCP vector needs to be constructed by extracting CCPs from the images. It has the general form:

\begin{equation}\label{equ:ccp-vector}
    \bm{w} = \left[ {\begin{array}{*{20}{c}}
                    {{\bm{w_1}}} \\
                    \vdots       \\
                    {{\bm{w_m}}}
                \end{array}} \right] = \left[ {\begin{array}{*{20}{c}}
                    {\bm{p}_1^1} & \ldots & {\bm{p}_1^n} \\
                    \vdots       & \ddots & \vdots       \\
                    {\bm{p}_m^1} & \ldots & {\bm{p}_m^n}
                \end{array}} \right]
\end{equation}
where $\bm{w}$ is the CCP vector, $m$ is the number of images, $n$ is the number of CCPs in each image, and $\bm{p}$ is the pixel vector that contains colour values, either intensity or chromaticity, which is calculated by averaging each colour patch.

In the camera response calibration step, the optimal iCRF is determined to be the one that produces the lowest colour distortions among the CCPs gathered across the calibration images:

\begin{equation}\label{equ:camera-response-calibration}
    \mathop{\arg\min}_{\bm{\theta}} \mathcal{J}\left(f_{\bm{\theta}}^{-1} \left(\bm{w_u}\right)\right) \\
\end{equation}
where $\bm{w_u}$ denotes CCPs collected from calibration images.

In the next response linearisation step, the images under examination are colour intensity linearised to $\bm{v'}$ by the iCRF determined in the first step. This process is denoted by Eq. \ref{equ:camera-reponse-correction}.

In the final colour matching step, colour intensity and chromaticity of the colour-linearised images produced in the previous step are matched to each other based on the CCP vector extracted from the test images which is denoted by $\bm{w_v}$. Linear transformation coefficients $\bm{\alpha}$ and $\bm{\beta}$ are firstly estimated from CCP vector $\bm{w_v}$ through a regression process by minimising the colour deviations between images as in Eq. \ref{equ:colour-matching-linear-regression}. Two is the minimum number of CCPs needed for this estimation since two points determine a line.

\begin{equation}\label{equ:colour-matching-linear-regression}
    \mathop{\arg\min}_{\bm{\alpha}, \bm{\beta}} = \mathcal{J}\left(\left[ {\begin{array}{*{20}{c}}
            {\bm{\alpha}_1 \bm{p}_1^1 + \bm{\beta}_1} & \ldots & {\bm{\alpha}_1 \bm{p}_1^n + \bm{\beta}_1} \\
            \vdots                 & \ddots & \vdots                 \\
            {\bm{\alpha}_m \bm{p}_m^1 + \bm{\beta}_m} & \ldots & {\bm{\alpha}_m \bm{p}_m^n + \bm{\beta}_m}
        \end{array}} \right]\right)
\end{equation}

The calculated coefficients are then used to perform linear colour matching on the colour-linearised images:

\begin{equation}\label{equ:colour-matching}
    \bm{v''} = \bm{\alpha}\bm{v'}+\bm{\beta} = \left[ {\begin{array}{*{20}{c}}
                    {\bm{\alpha}_1 \bm{v'}_1+\bm{\beta}_1} \\
                    \vdots         \\
                    {\bm{\alpha}_m \bm{v'}_m+\bm{\beta}_m}
                \end{array}} \right]
\end{equation}

In the rest of this section, the proposed camera response calibration is firstly detailed, and followed by an explanation of the proposed colour matching algorithms.

\subsection{Camera response calibration}

In this paper, camera response calibration is seen as the process of estimating the nonlinearity in colour intensity. Two calibration approaches are proposed: 
\begin{enumerate}
    \item \emph{Selection} of the optimal CRF from the DoRF by exhaustive search.
    \item Calculation of the optimal CRF model parameters by machine learning based \emph{Optimisation}.
\end{enumerate}
The EMoR is used as the CRF representation model in the Optimisation approach.

\begin{figure}[htbp] 
    \centering
    \subfloat[BoLD of raw images taken by a Cannon 60D camera. The mean distance curve (in green) is almost symmetric.]{\includegraphics[width=1.0\linewidth]{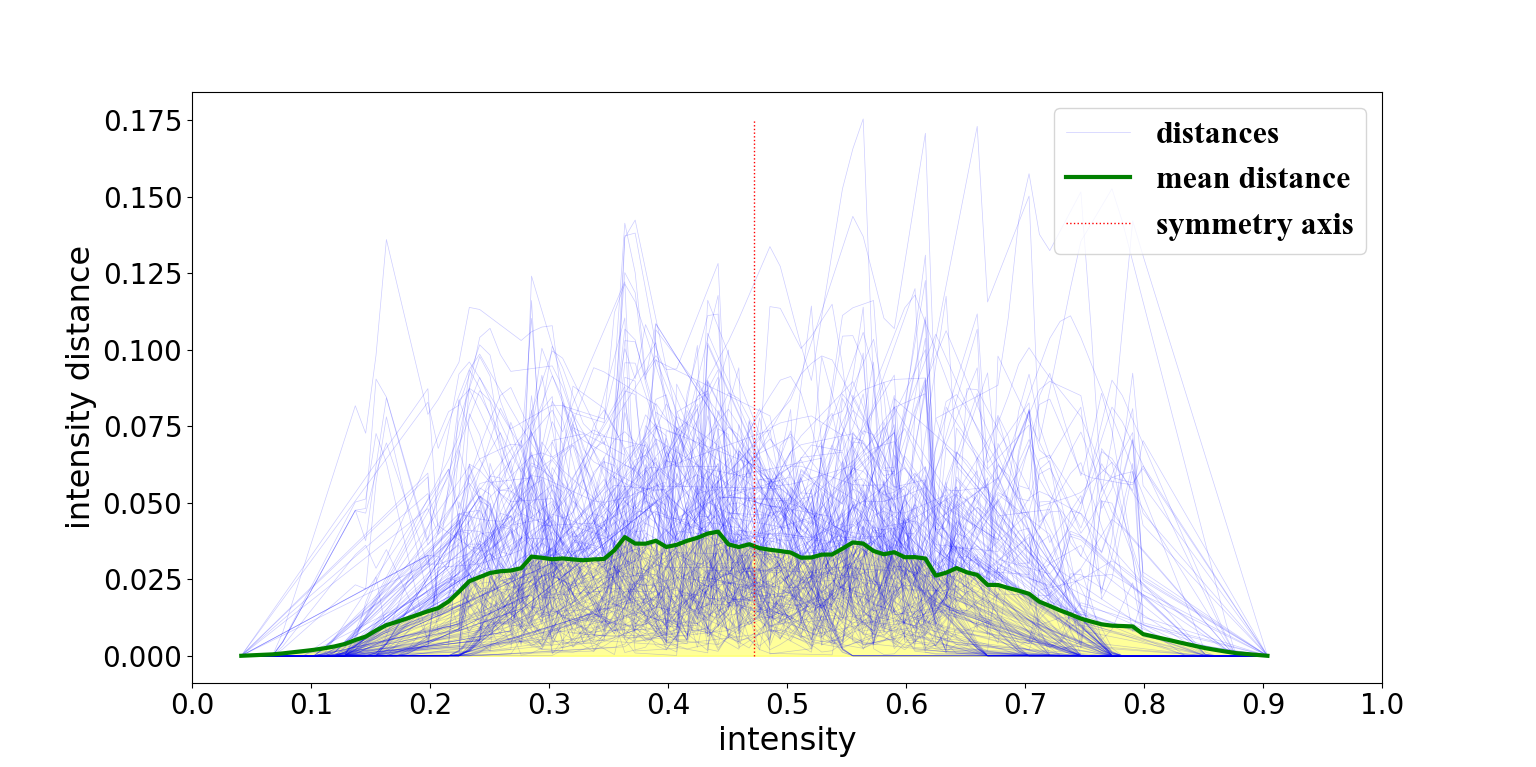}}
    \hfil
    \subfloat[BoLD of final images taken by a Hisense H10 smartphone. The mean distance curve (in green) is clearly asymmetric.]{\includegraphics[width=1.0\linewidth]{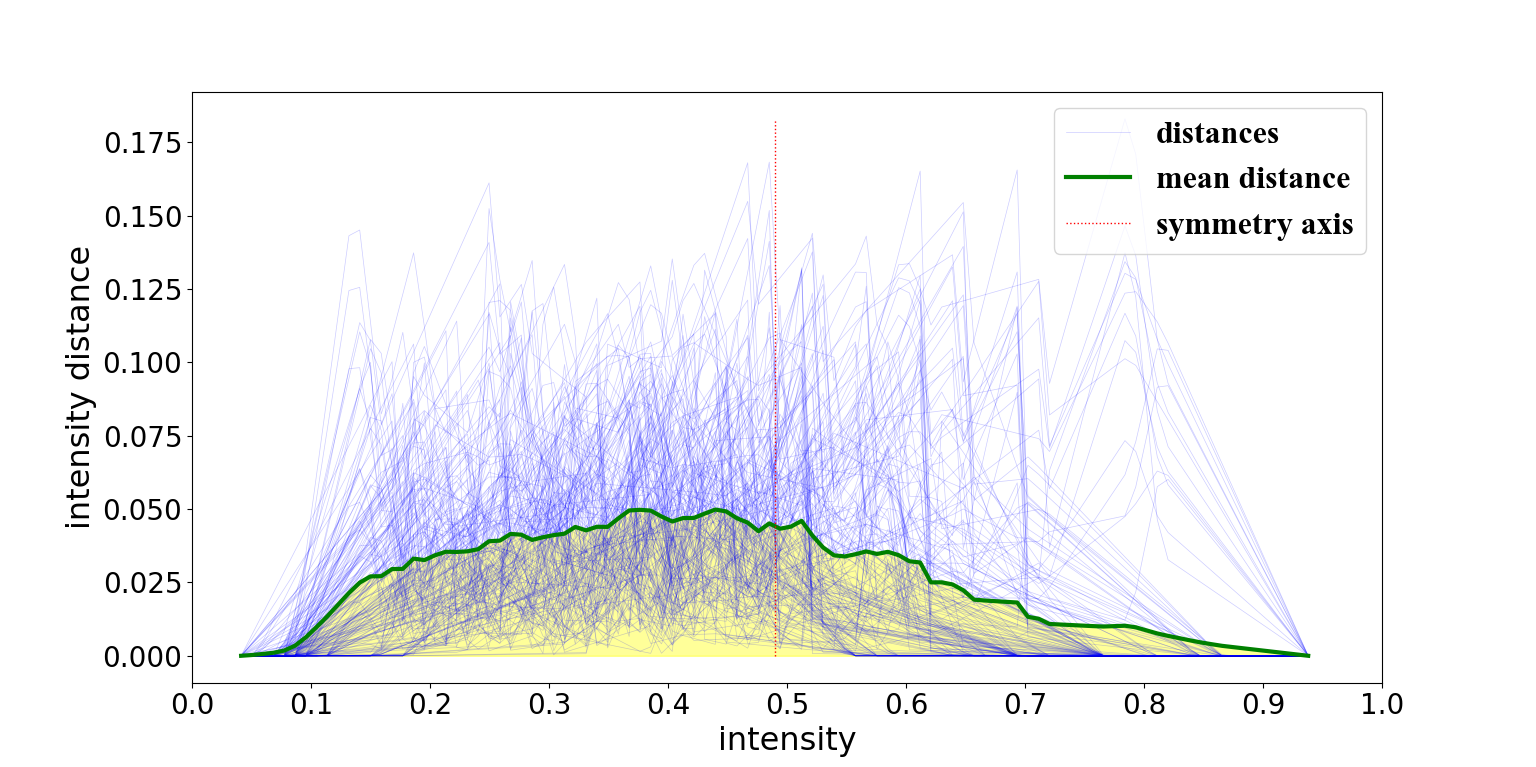}}
    \caption{Visual explanation of the BoLD feature. 20 images were taken by each of these cameras. The images were taken under various uniform illuminations. 24 CCPs were extracted from each of these images. The CCPs were sorted based on their intensity values. The blue lines are rendered from the calculated $\bm{d}$. The green curve was drawn from the mean intensity distance $\bm{\bar d}$. The area under the $\bm{\bar d}$ curve is marked yellow. And the red dashed line denotes the symmetry axis of $\bm{\bar d}$.}
    \label{fig:distance-visualisation}
\end{figure}

Intensity distance is the intensity offsets between CCPs. The intensity distances calculated from RAW images taken by a digital camera (i.e., images with linear CRF) is visualised in Fig. \ref{fig:distance-visualisation}(a). It can be seen that they are almost symmetric along the x-axis (i.e., w.r.t intensity of CCPs) when the applied CRF is linear. This is mainly due to the homogeneous and normal noise distribution across the colour intensity range during imaging. Images taken by another camera with a nonlinear CRF led to asymmetric intensity distances as shown in Fig. \ref{fig:distance-visualisation}(b).

Based on this finding, both of the proposed approaches evaluate candidate iCRFs by a novel balance-of-linear-distances (BoLD) feature which assumes a balanced intensity distances from images produced with a linear CRF. The BoLD feature is determined by the skewness of the mean intensity distance curve. Based on this BoLD feature, a BoLD value is proposed by combining the normalised feature with the area under the mean intensity distance curve. Calculation of this value does not require a knowledge of true colour values in the image and works in an unsupervised process.

In order to calculate this BoLD value, colour intensity values of the CCPs are firstly extracted from the images used for calibration and encapsulated in a two-dimensional CCP vector similar to that in Eq. \ref{equ:ccp-vector}. The rows of the vector represent the different images used for calibration while the columns represent varied intensities of the CCPs. The columns of the CCP vector are sorted according to their mean values. Then, colour intensity values in the CCP vector are converted into irradiance values and aligned with each other based on the first and last columns in the sorted CCP vector:

\begin{equation}\label{equ:colour-matching-linear-regression}
    \mathop{\arg\min}_{\bm{\alpha}, \bm{\beta}} = \mathcal{J}\left(\left[ {\begin{array}{*{20}{c}}
            {\bm{\alpha}_1 \bm{p}_1^1 + \bm{\beta}_1} & {\bm{\alpha}_1 \bm{p}_1^n + \bm{\beta}_1} \\
            \vdots            & \vdots            \\
            {\bm{\alpha}_m \bm{p}_m^1 + \bm{\beta}_m} & {\bm{\alpha}_m \bm{p}_m^n + \bm{\beta}_m}
        \end{array}} \right]\right)
\end{equation}

\begin{equation}\label{equ:processed-CCP-vector}
    \bm{w_u^\prime} = \left[ {\begin{array}{*{20}{c}}
                    {\bm{\alpha}_1 f_{\bm{\theta}}^{ - 1}\left(\bm{p}_1^1\right)+\bm{\beta}_1} & \ldots & {\bm{\alpha}_1 f_{\bm{\theta}}^{ - 1}\left(\bm{p}_1^n\right)+\bm{\beta}_1} \\
                    \vdots                               & \ddots & \vdots                               \\
                    {\bm{\alpha}_m f_{\bm{\theta}}^{ - 1}\left(\bm{p}_m^1\right)+\bm{\beta}_m} & \ldots & {\bm{\alpha}_m f_{\bm{\theta}}^{ - 1}\left(\bm{p}_m^n\right)+\bm{\beta}_m}
                \end{array}} \right]
\end{equation}
where $\bm{w_u^\prime}$ is the colour-aligned CCP vector.

The two-dimensional mean CCP vector $\bar W$ is constructed by joining $m$ one-dimensional mean CCP vectors $\bm{\bar w}$ themselves calculated by averaging colour-aligned CCPs along the calibration images:

\begin{equation}\label{equ:caculate-mean-cp}
    \bm{\bar w} = \frac{{\sum_{i=1}^m {f_{\bm{\theta}}^{-1}\left(\left(\bm{{w_u^\prime}}\right)_i \right)} }}{m}
\end{equation}

\begin{equation}\label{equ:compose-mean-cp-matrix}
    \bar W = \left. {\left[ {\begin{array}{*{20}{c}}
                    {\bm{\bar w}} \\
                    \vdots        \\
                    {\bm{\bar w}}
                \end{array}} \right]} \right\}m
\end{equation}

Intensity distances are calculated as the element-wise distance between the colour-aligned CCP vector $\bm{w_u^\prime}$ and mean CCP vector $\bar W$:

\begin{equation}\label{equ:intensity-distance}
    \bm{d} = \left| \bm{w_u^\prime}  - \bar W \right|
\end{equation}

And the mean intensity distance curve is produced by averaging $\bm{d}$ in the column direction:

\begin{equation}\label{equ:caculate-mean-intensity-distance}
    \bm{\bar d} = \frac{{\sum_{i=1}^m {\bm{d}_i} }}{m}
\end{equation}

Eventually, the BoLD value, denoted as $\mathcal{B}$, is calculated by the l2-norm of the normalised asymmetry coefficient and the weighted area under the $\bm{\bar d}$ curve:

\begin{gather}\label{equ:bold}
    \mathcal{B} = \left \|\eta - \lambda_1 \phi, \lambda_2 \mu \right \|_2 \\
    \eta = \frac{\sum \left ( \bm{x}-\bm{\overline{x}}\right )^3}{s\left (\frac{ \sum \left ( \bm{x}-\bm{\overline{x}}\right )^2}{s} \right )^\frac{3}{2}} \\
    \phi = \max\left(\bm{\bar d}\right) + \min\left(\bm{\bar d}\right) - 1 \\
    \mu = \sum \bm{\bar d}
\end{gather}
where $\eta$ is the asymmetry coefficient, $\phi$ is the normalisation term, $\mu$ denotes the area under the $\bm{\bar d}$ curve, $\bm{x}$ and $\bm{\bar x}$ are the uniform sampled values on $\bm{d}$ and mean of the samplings. The number of samplings $s$ was empirically selected to be 100. The normalisation term is included and weighted by $\lambda_1$ to minimise the effect due to skewed samplings in the colour intensity range, i.e., intensities of CCPs in the images for calibration are not normally distributed in $\left[0, 1.0\right]$. The area under the curve is properly weighted by $\lambda_2$ to adjust correction magnitude.

The Selection approach directly chooses the optimal iCRF from the DoRF. The optimal iCRF is selected by exhaustively searching the 201 CRFs in the DoRF and choosing the best one with minimum BoLD value as the cost.

In addition to the BoLD value, smoothness terms are added into the cost function of the Optimisation approach for preventing discontinuous iCRF. Monotonicity in the CRF is observed from CRFs in the DoRF. Thus, the iCRF is also monotonic, and the monotonicity is included as a constraint in the cost function for minimising ambiguity. The optimal EMoR model parameters $\bm{\theta}$ is calculated as:

\begin{gather}\label{equ:optimisation-cost}
    {\bm{\theta}} = \mathop {\arg \min }\limits_\theta \left\| \mathcal{B},{\psi _1} \left| {{f_{\bm{\theta}}^{ - 1}}^{\prime \prime} } \right|, {\psi _2} \sqrt {\frac{{{{\sum^M {\left( {{f_{\bm{\theta}}^{ - 1}}^{\prime \prime}  - {\overline {f_{\bm{\theta}}^{ - 1}}^{\prime \prime} } } \right)} }^2}}}{M}} \right\|_2^2 \nonumber\\
    {\rm{\ subject\ to\ }}{f_{\bm{\theta}}^{ - 1}}^{\prime}  > 0
\end{gather}
where the second term is the micro-smoothness restriction that prevents discontinuous iCRFs and is weighted by $\psi_1$, the third term is the macro-smoothness restriction that maintains the overall shape of iCRF and is weighted by $\psi_2$, $M$ is the number of homogeneous macro-samplings on the candidate iCRF and was set to 10, the monotonicity is ensured by a positive first-derived iCRF.

Eventually, the Optimisation approach reconstructs the iCRF from the calculated EMoR parameters with a form similar to that of Eq. \ref{equ:dorf}:

\begin{equation}\label{equ:icrf}
    f^{-1} = H_0^{-1}+\bm{\theta}^T H_{1:k}^{-1}
\end{equation}
where $H_{0:k}^{-1}$ is the mean and first $k$ eigenvectors of all iCRFs in the DoRF.

\subsection{Pixel-wise linear colour intensity and chromaticity matchings}
\label{subsection:linear-matchings}

\begin{algorithm}[htbp]
    \caption{Pixel-wise linear colour matching algorithm}
    \label{alg:li-colour-matching}
    \begin{algorithmic}[1]
        \Require
        $image1$ is the image to be matched and
        $image2$ is the image to be corrected;
        \Ensure
        $image2$;
        \State $R_1$, $G_1$, $B_1$ $\leftarrow$ $R$, $G$, $B$ values of the CCPs in $image1$;
        \State $R_2$, $G_2$, $B_2$ $\leftarrow$ $R$, $G$, $B$ values of the CCPs in $image2$;
        \State $I_1 \leftarrow R_1+G_1+B_1$;
        \State $I_2 \leftarrow R_2+G_2+B_2$;
        \State $\alpha_I$, $\beta_I$ $\leftarrow$ linear regression on $\left\{I_1,I_2\right\}$;
        \State $\alpha_r$, $\beta_r \leftarrow$ linear regression on $\left\{\frac{R_1}{I_1},\frac{R_2}{I_2}\right\}$;
        \State $\alpha_b$, $\beta_b \leftarrow$ linear regression on $\left\{\frac{B_1}{I_1},\frac{B_2}{I_2}\right\}$;
        \State $image2$ $\leftarrow$ $I_2$ scaled by $\alpha_I$ and offset by $\beta_I$;
        \State $image2$ $\leftarrow$ $\frac{R_2}{I_2}$ scaled by $\alpha_r$ and offset by $\beta_r$;
        \State $image2$ $\leftarrow$ $\frac{B_2}{I_2}$ scaled by $\alpha_b$ and offset by $\beta_b$;
        \\
        \Return $image2$;
    \end{algorithmic}
\end{algorithm}

After the optimal iCRF has been calibrated, images taken with the same camera can be transformed into a response-linear space by interpolation on the estimated iCRF as shown in Eq. \ref{equ:camera-reponse-correction}. Once linearisation is complete, the colour intensity and chromaticity of any two such images are linearly matched. The colour intensity and chromaticity of a second image is linearly matched to that of the first using Algorithm \ref{alg:li-colour-matching}. Note that the $\alpha$ and $\beta$ coefficients in this algorithm is the same to those in the previous sections except that they are in general forms in the previous sections and are specific, i.e., for intensity and chromaticity, in this algorithm. The algorithm to independently modify the colour intensity and chromaticity without affecting the other component is summarized in Algorithm \ref{alg:colour-modification}.

\begin{algorithm}[htbp]
    \caption{Pixel-wise independent colour modification algorithm}
    \label{alg:colour-modification}
    \begin{algorithmic}[1]
        \Require
        $image1$ is the image to be independently colour intensity and RB chromaticity modified. $\alpha_I$ and $\beta_I$ are the colour intensity scaling and offset coefficients. $\alpha_r$ and $\alpha_b$ are the scaling coefficients for red and blue chromaticity, respectively. $\beta_r$ and $\beta_b$ are the offset coefficients for red and blue chromaticity, respectively;
        \Ensure
        $image2$;
        \State $R^\prime$, $G^\prime$, $B^\prime$ $\leftarrow$ $R$, $G$, $B$ pixel values in $image1$;
        \State $R^\prime \leftarrow \alpha_I R^\prime$;
        \State $G^\prime \leftarrow \alpha_I G^\prime$;
        \State $B^\prime \leftarrow \alpha_I B^\prime$;
        \State $R \leftarrow \frac{\left(R^\prime+G^\prime+B^\prime+\beta_I \right) R^\prime}{R^\prime+G^\prime+B^\prime}$;
        \State $B \leftarrow \frac{\left(R^\prime+G^\prime+B^\prime+\beta_I \right) B^\prime}{R^\prime+G^\prime+B^\prime}$;
        \State $G \leftarrow R^\prime+G^\prime+B^\prime+\beta_I-R-B$;
        \State $R^\prime \leftarrow \alpha_r R$;
        \State $B^\prime \leftarrow \alpha_b B$;
        \State $G^\prime \leftarrow R+G+B-\alpha_r R-\alpha_b B$;
        \State $R \leftarrow \beta_r \left(R^\prime+G^\prime+B^\prime\right)+R^\prime$;
        \State $B \leftarrow \beta_b \left(R^\prime+G^\prime+B^\prime\right)+B^\prime$;
        \State $G \leftarrow R^\prime+G^\prime+B^\prime-R-B$;
        \State $image2 \leftarrow R, G, B$;
        \\
        \Return $image2$;
    \end{algorithmic}
\end{algorithm}

\section{Experimental Setup}

This section details the experimental setup used to examine and test the proposed model. Firstly, the three image datasets used for the performance evaluations are introduced. Then, the evaluation procedure and metrics and comparison methods for evaluating are explained. Finally, the implementation details used are described.

\subsection{Image datasets}

\begin{figure}[htbp]
    \centering
    \subfloat[The created Belfast dataset]{\includegraphics[width=1.0\linewidth]{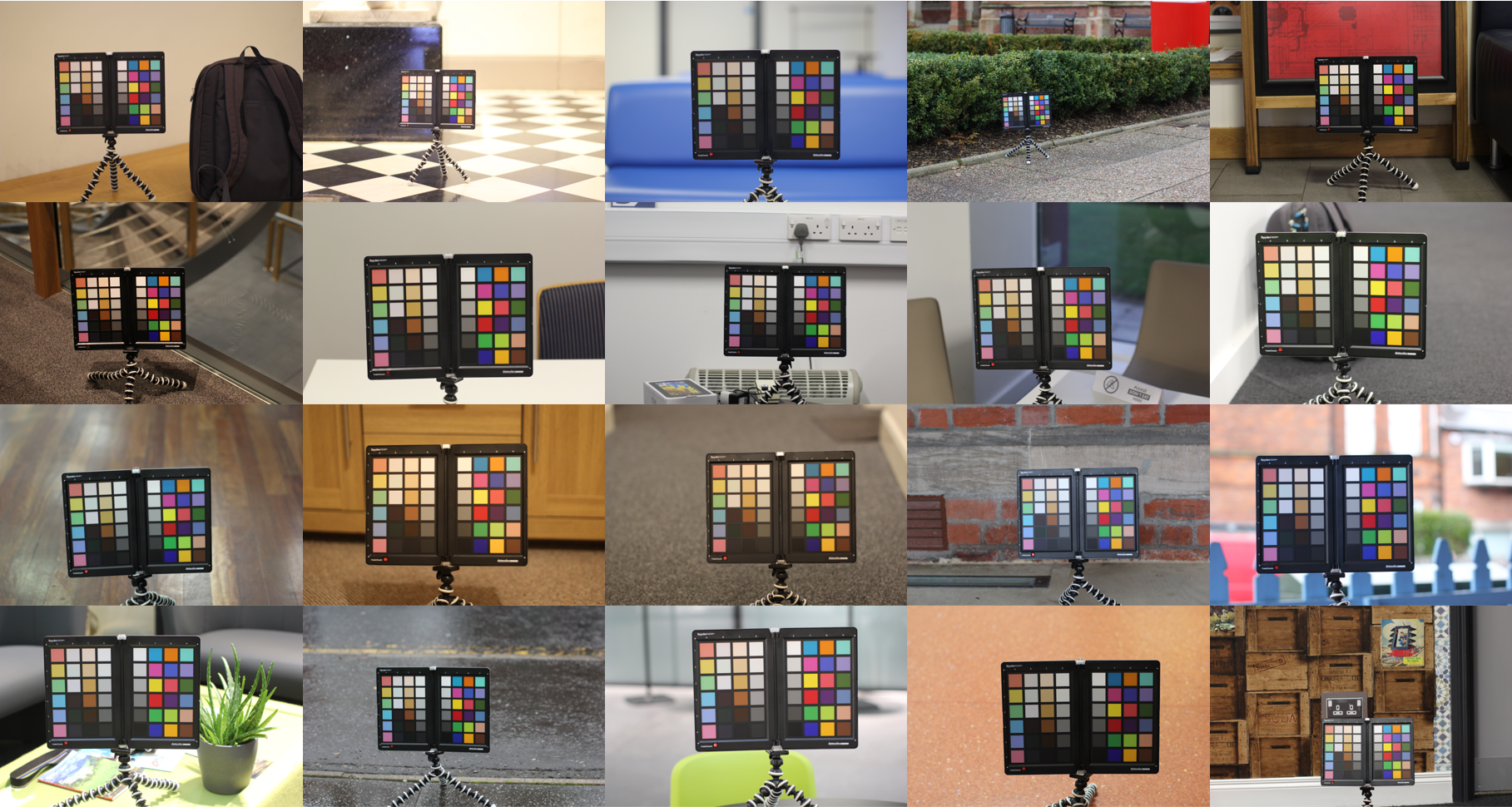}}
    \\
    \subfloat[The modified Middlebury \cite{chakrabarti2009empirical} dataset]{\includegraphics[width=0.95\linewidth]{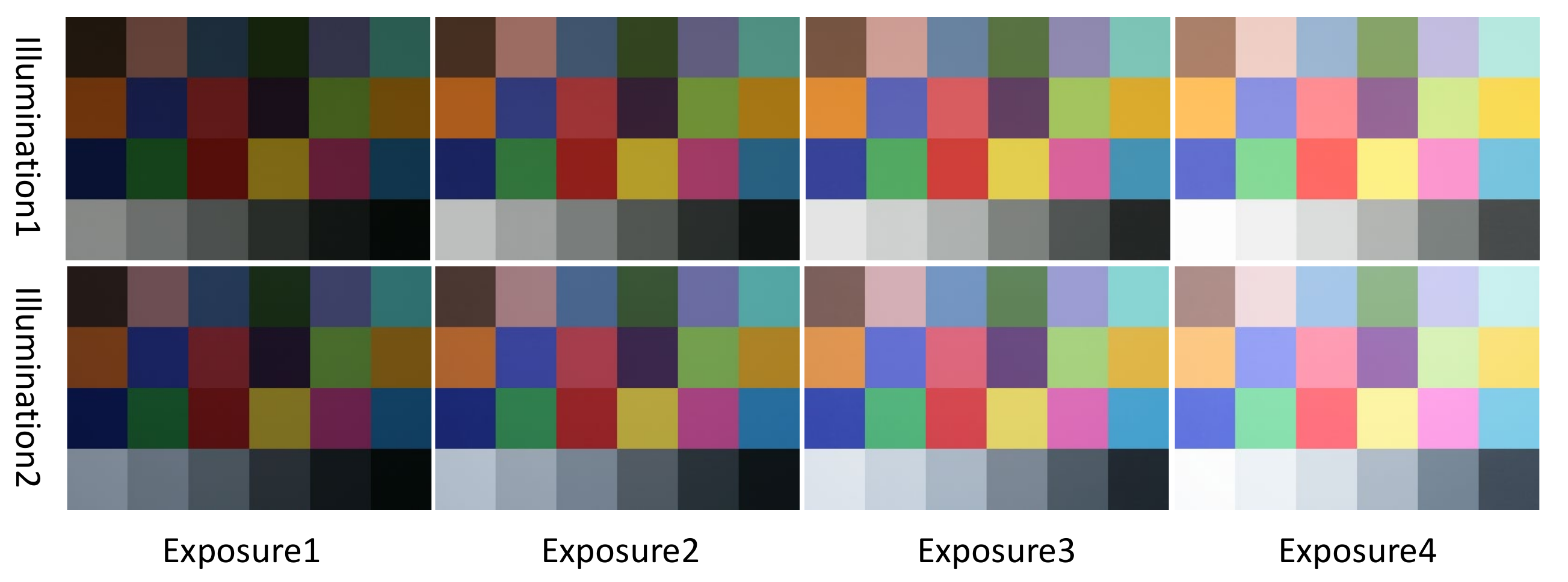}}
    \\
    \subfloat[The rendered Gehler-Shi \cite{Gehler2008, Afifi2019} dataset]{\includegraphics[width=0.9\linewidth]{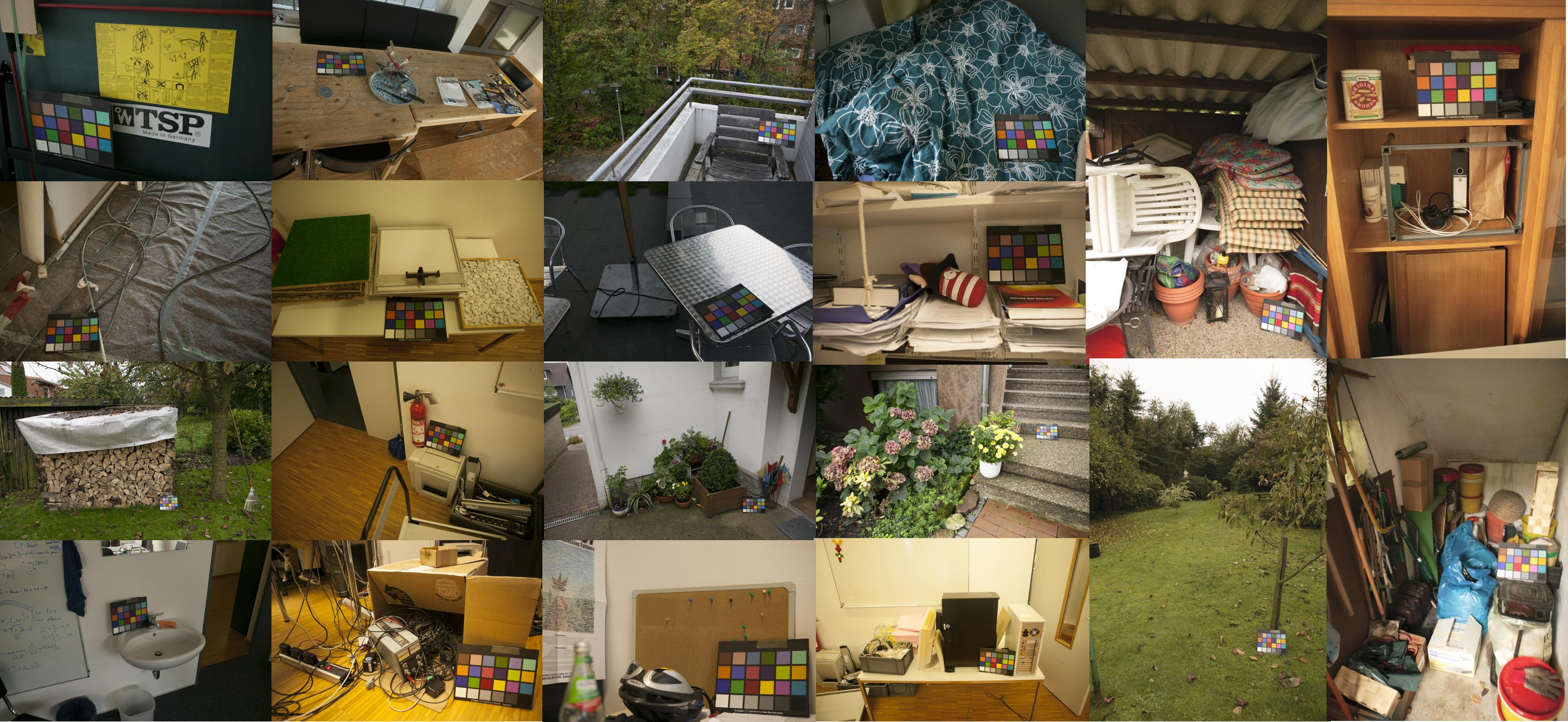}}
    \caption{Sample images taken by one of the camera in the three datasets being used.}
    \label{fig:dataset-cc}
\end{figure}

Most existing image datasets for CC research \cite{Aytekin2017, Cheng2014, Hemrit2018} lack of images taken by different smartphone cameras. In order to thoroughly evaluate RCC performance of the proposed model, three image datasets were used for performing the validations and benchmarks. These datasets include a newly created image dataset and two image datasets modified and rendered from the Middlebury \cite{chakrabarti2009empirical} and Gehler-Shi \cite{Gehler2008, Afifi2019} datasets. Sample images of these three datasets are demonstrated in Fig. \ref{fig:dataset-cc}. 

The created image dataset, namely Belfast, contains 120 images taken by six different cameras. Images by each camera were taken under 20 various and uniform illuminations. The six cameras used were rear cameras of a Samsung Galaxy S7 (Samsung, South Korea), a Hisense H10 (Hisense, China), an iPhone6P (Apple, USA), and an iPad (Apple, USA), and two digital single lens reflex (DSLR) cameras: a Canon 60D (Canon, Japan) and an Olympus E-M10II (Olympus, Japan). Among the 20 illuminations, 16 of them were indoor and the rest were outdoor. Two 24-patch colour charts in a Spyder SCK100 (Datacolor, USA) ColorChecker were included in a common everyday-life scene in each image. The colour charts in the images were all placed near the centre of the imaging field and perpendicular to the line of camera sight. The images were captured without High Dynamic Range (HDR) and using auto-WB. 

The modified Middlebury \cite{chakrabarti2009empirical} dataset contains a total of 112 images with eight images taken by each of the 14 selected DSLR cameras from the original dataset. These cameras were selected due to their higher cross-channel response uniformity. Images were taken under two illuminations and four exposures for each of the 14 camera models. Each image is a Macbeth colour chart without any other content. This dataset can imitate scientific imaging scenes. 

A total of 336 images were rendered and selected from the Gehler-Shi \cite{Gehler2008, Afifi2019} dataset. These images were generated from 14 different profiles of WB and image post-processing procedures. A Macbeth colour chart was placed in common everyday-life scenes in the images. Four images with varied illuminations were selected to be the calibration images for each profile.

The CCP locations in the images (48 CCPs for each image in the created Belfast dataset and 24 CCPs for each image in the other two datasets) were carefully labelled by utilising a custom-developed Python script so that the CCPs can be aligned and compared with each other across different images. The reason for including colour charts in the images is for the ease of CCP extraction and comparison. The true colour values of the CCPs on colour charts such as reflective spectrum, white point, and CIE colour values were \emph{not} included in the labels since RCC evaluates colour alignment between measurements rather than that between measurements and a standard.

\subsection{Evaluation procedure}

All the three datasets were divided into two subsets, one used for calibration, and the other used for testing and evaluation. For the Belfast and modified Middlebury \cite{chakrabarti2009empirical} datasets, the number of images taken by each camera is evenly allocated to the calibration and testing subsets, i.e., ten calibration and ten testing images for each camera in the Belfast dataset and four calibration and four testing images for each camera in the modified Middlebury \cite{chakrabarti2009empirical} dataset. While it is four calibration images and 20 testing images for each profile in the rendered Gehler-Shi \cite{Gehler2008, Afifi2019} dataset. For the Belfast dataset, only the Macbeth colour chart (24 CCPs) was used for calibration, while both of the colour charts (48 CCPs) were used for performance evaluation. For the other two datasets, both the calibration and performance evaluation used 24 CCPs.

To produce the colour-aligned images for evaluation, the CCP vector was firstly extracted from the colour patches of the colour charts appeared in the images used for calibration. The iCRF of each camera was calibrated by using the extracted CCPs. Then, the calibrated iCRF of each camera was used to linearise every test image taken by that camera. The colour-linearised image was colour matched to the image to be compared to produce the colour-aligned image. Eventually, the CCPs were extracted from the colour-aligned images and quantitatively evaluated for the RCC performance.

\subsection{Evaluation metrics}
\label{subsection:eva-metrics}

The root-mean-square error (RMSE) \cite{Lin2004, Sharma2020, Grossberg2004, Li2017, Gehler2008}, recovery angular error (RAE) \cite{Karaimer2018, Finlayson2014, Gehler2008, Laakom2019, VandeWeijer2007}, and $\Delta$E 2000 \cite{Afifi2019, Afifi2020} are common metrics to quantify colour difference. The RMSE is an ideal metric for quantifying colour intensity difference as it measures the Euclidean distance between two compared CCPs. However, the RMSE is inefficient when measuring chromaticity difference since chromaticity is the ratio of a colour channel to the intensity. The RAE measures the angular difference between two pixels or colours as directed vectors and is seldom affected by colour intensity. Thus, it is used for estimating the chromaticity difference between the compared CCPs. And the $\Delta$E 2000 is used to quantify colour difference in visual perception.

A smaller RMSE, RAE, or $\Delta$E 2000 value indicates a better result. A $0$ value illustrates identical colour of the compared CCPs measured by that metric.

\subsection{Comparison methods}

The handshake comparison strategy was used to evaluate the RCC performance of an image collection. Each item, i.e., either an image or a camera, will have a chance to be compared with every other item in the comparison collection. The number of comparisons is therefore $\frac{1}{2}N\left(N-1\right)$ where $N$ is the number of items in the comparison collection, e.g., number of images or camera models \cite{Zhao2018}.

The handshake comparison results, either in RMSE, RAE, or $\Delta$E 2000, of images taken by each camera, $\bm{e}$, known as \emph{single camera} RCC performance, were firstly produced. For \emph{cross camera} RCC performance evaluation, different cameras were also handshake-compared. The comparison results from the handshake comparisons were collected into a result vector $\bm{z}$ for statistical analysis:

\begin{equation}\label{equ:comparison-result-vector}
    \bm{z} = \left. {\left[ {
                \begin{array}{*{20}{c}}
                    {\bm{e}_0} \\
                    \vdots  \\
                    {\bm{e}_{-1}}
                \end{array}} \right]} \right\}\frac{1}{4}m\left(m-1\right)c\left(c-1\right)
\end{equation}
where $m$ is the number of images by each camera, and $c$ is the number of camera models to be compared.

The \emph{Median} was calculated from the handshake comparison results $\bm{z}$. It was selected to be the best overall RCC performance indicator as $\bm{z}$ is \textbf{not} normally distributed, and impacts due to outliers can be minimised.

\subsection{Implementation details}
 
The linear interpolation provided in NumPy \cite{oliphant2006guide} was used to interpolate the iCRF. SciPy \cite{bressert2012scipy} was selected for conducting the linear regression in the proposed colour matching algorithms. Tensorflow \cite{Abadi2016} was used as the machine learning framework for implementing the Optimisation approach. Since the solution space is non-convex, 50 random initialisations were applied for each optimisation. The Adam optimiser \cite{Kingma2014} was applied for executing the optimisation. An exponential decayed learning rate with an initial value of 0.5, decay step of 1000, and decay rate of 0.9 was adopted to the optimiser. The optimisation was performed for a maximum of 600 epochs or terminated by a delta-cost less than $10^{-3}$.

All the processing and evaluations were performed on a laptop computer with a 2.6-GHz Intel Core i7 processor and a 16-GB memory. To accelerate the optimisation process, a NVIDIA GeForce RTX 2060 GPU was employed.

\section{Experimental Results}

This section demonstrates the outcomes of the experiments and tests that reflect performance of the proposed colour alignment model. Initially, the overall RCC performance of the proposed model with utilising the BoLD feature and Selection calibration approach (namely, BoLD-alignment) is demonstrated. Then, the key model parameters and components are evaluated and ablation studied. Finally, performance benchmarks that compare the proposed model with the other popular and state-of-the-art methods are illustrated.

\subsection{The overall RCC performance}

\begin{figure*}[htbp]
    \centering
    \subfloat{\includegraphics[width=0.36\linewidth,valign=t]{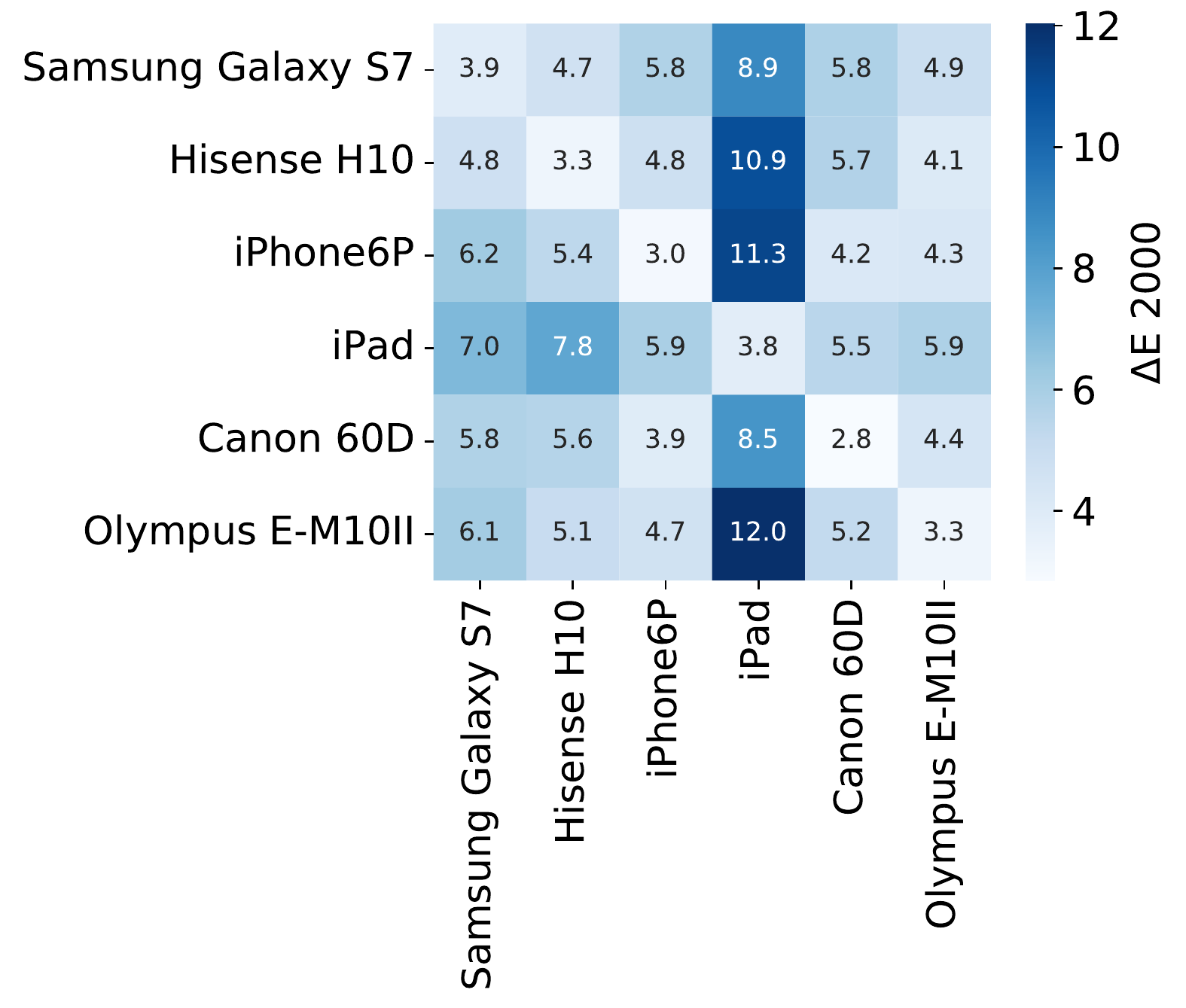}}
    \hspace{0.2cm}
    \subfloat{\includegraphics[width=0.34\linewidth,valign=t]{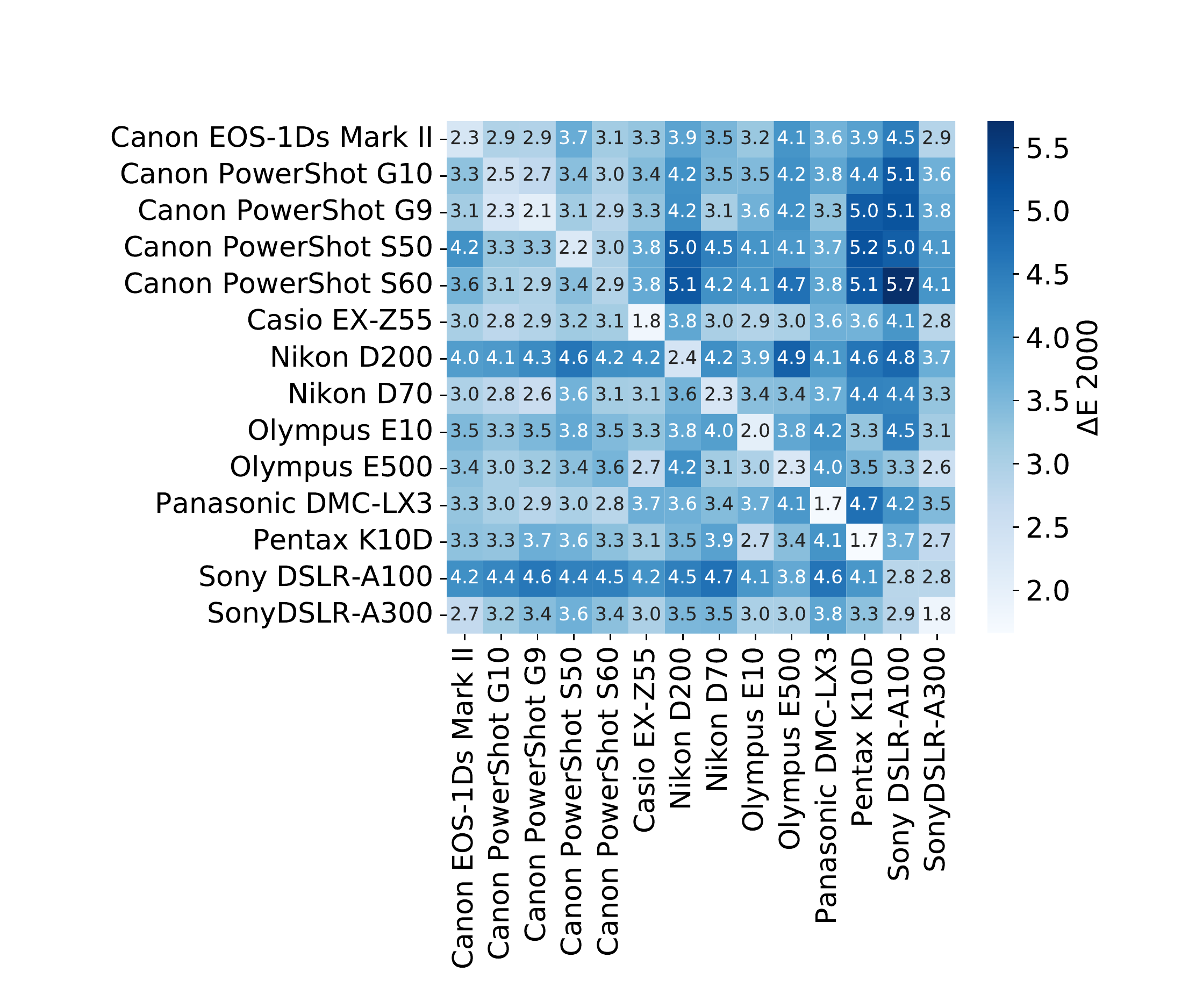}}
    \hspace{0.4cm}
    \subfloat{\includegraphics[width=0.25\linewidth,valign=t]{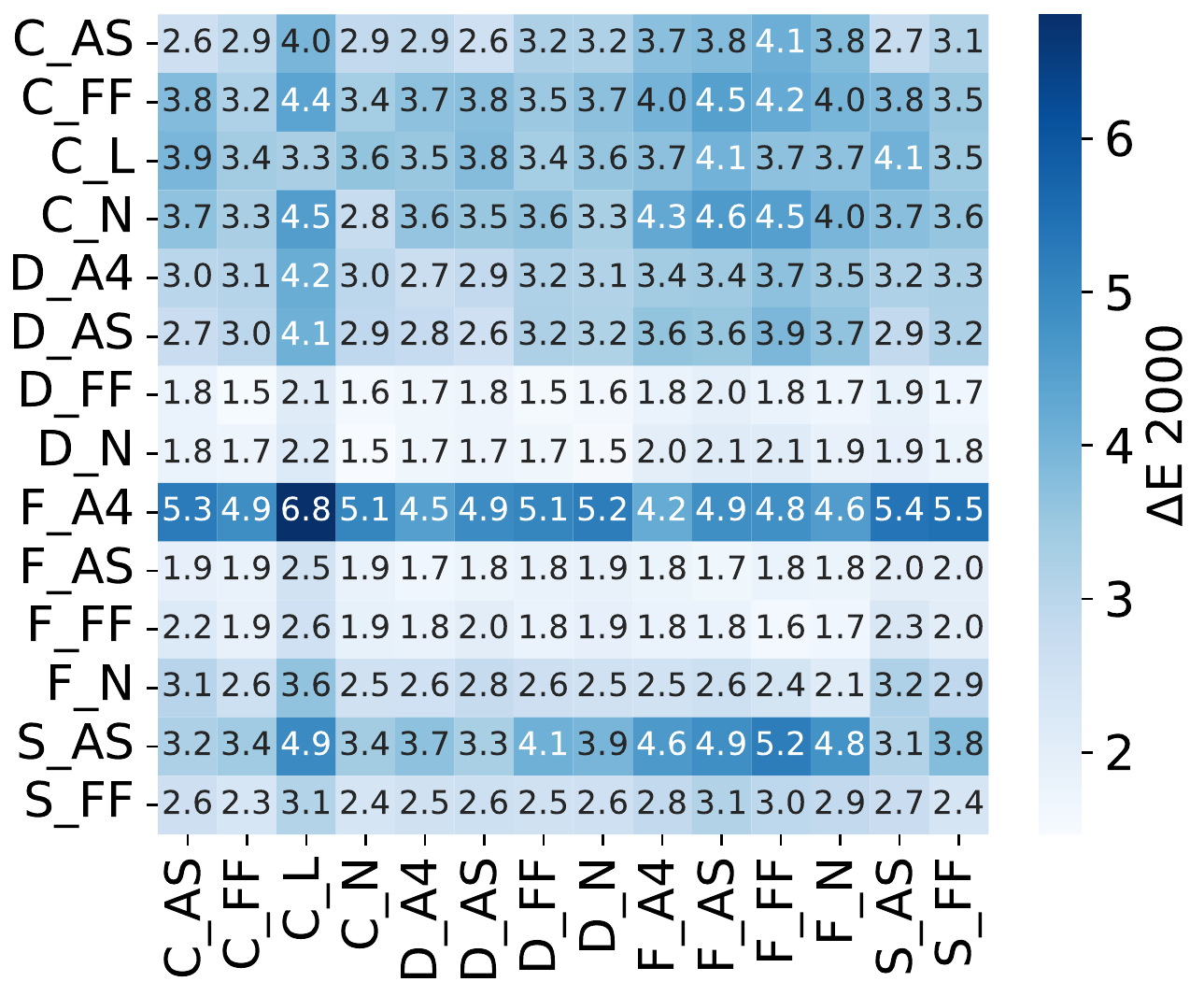}}
    \caption{Heatmap visualisation of the single and cross camera RCC performance on the Belfast (left), modified Middlebury \cite{chakrabarti2009empirical} (middle), and rendered Gehler-Shi \cite{Gehler2008, Afifi2019} (right) datasets. The BoLD-alignment performance was measured by $\Delta$E 2000.} 
    \label{fig:rcc-performance}
\end{figure*}

Fig. \ref{fig:rcc-performance} demonstrates the heatmap visualisations of the handshake comparison results. NoCI, NoCCP-CA, and NoCCP-CM used for producing these results were four, 24, and two, respectively. Cross-validation was applied for performance evaluation. The RCC performance of the proposed BoLD-alignment were evaluated in terms of the $\Delta$E 2000 on all the three image datasets. 

The diagonal values of each heatmap reflect single camera colour alignment performance, while the rests represent cross camera performance. A homogeneous heatmap with minimum error is ideal, which indicates close single and cross camera RCC performance. The results reveal that the handshake comparisons are not symmetric, i.e., colour alignment from image A to B does not equal to that from image B to A, yet, are symmetrically correlated. We can also see from the figure that there are cameras consistently generated significant larger errors, e.g., iPad and F\_A4 in the Belfast and rendered Gehler-Shi \cite{Gehler2008, Afifi2019} dataset. This is probably due to inaccurate response calibration or more sophisticated image post-processing applied on images taken by these cameras.

\subsection{Model parameters}

In this section, the created Belfast dataset was used for all the evaluations. A total of four model parameters were investigated.

\subsubsection{Number of calibration images (NoCI)}

\begin{figure*}[htbp]
    \centering
    \includegraphics[width=0.9\linewidth]{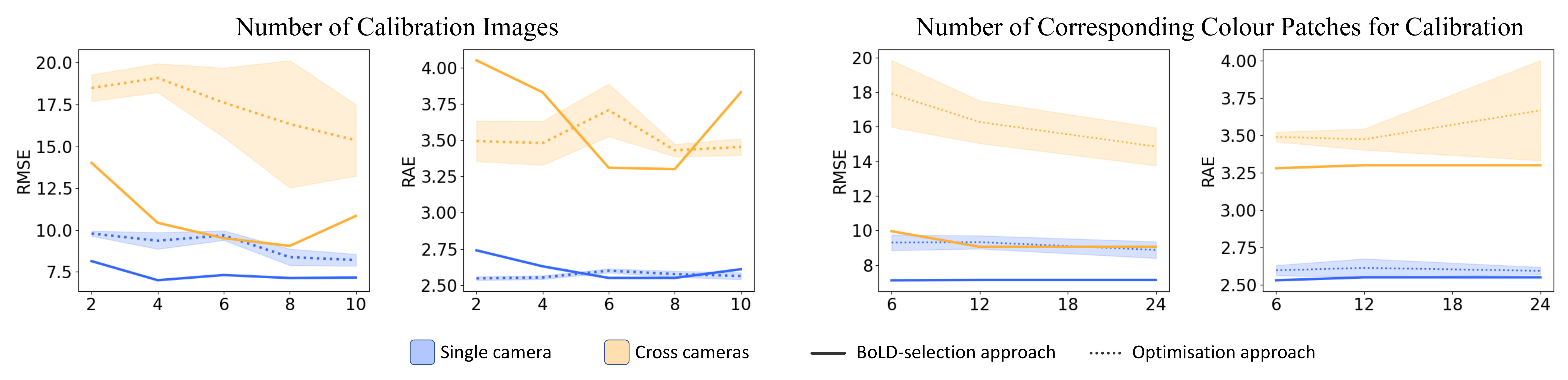}
    \caption{Single (blue curves) and cross (orange curves) RCC performance of the Optimisation (dashed curves) and Selection (solid curves) calibration approaches with applying different NoCIs (2, 4, 6, 8, and 10) and NoCCP-CAs (6, 12, and 24). The grey area indicates the standard derivation (SD, n=3) by the Optimisation approach.}
    \label{fig:noci}
\end{figure*}

The minimum NoCI required for the proposed calibration to perform is an important factor to be considered as the NoCI needed to be prepared by a user should be as low as possible, while still delivering satisfactorily accurate. The results demonstrated in Fig. \ref{fig:noci} were produced from both of the Optimisation and Selection calibration approaches while applying the process calibrated with a different NoCIs (i.e., 2, 4, 6, 8, and 10) but keeping NoCCP-CA fixed at 12 and NoCCP-CM fixed at 24. It can be observed from these results that the Selection approach outperformed the Optimisation approach for both single and cross camera RCC performance in terms of the RMSE. In addition, the single and cross camera RCC performance in terms of the RMSE produced by the Selection approach were closer to each other compared to those produced by the Optimisation approach. While the RAE performance did not show significant difference between the Selection and Optimisation approaches, nor between the single and cross camera RCC performance. In general, a larger NoCI helped enhance the overall camera colour calibration performance in terms of the RMSE yet has no effect on the performance in terms of the RAE. In terms of the RMSE, four was the minimum NoCI for the Selection calibration approach to perform satisfactorily, while it is over ten for the Optimisation approach. This indicates that the Selection approach is more applicable because of the smaller NoCI needed.

\subsubsection{Number of corresponding colour patches for calibration (NoCCP-CA)}

NoCCP-CA is also an important parameter to be investigated as obtaining a CCP is expensive in consumer-oriented imaging applications. The results were produced from the proposed model while applying varied NoCCP-CA (i.e., 6, 12, and 24) yet keeping NoCI fixed at eight and NoCCP-CM fixed at 24. Fig. \ref{fig:noci} presents that the Selection approach produced better overall single and cross camera RCC performance in terms of the RMSE and the RAE. Cross camera RCC performance in terms of the RMSE of the Optimisation approach improved as the NoCCP-CA increased. The performance indicators remained stable across all the NoCCP-CAs tested. This suggests that the Selection approach was the better choice due to the superior and more stable RCC performance. The smallest NoCCP-CA tested (n=6) was good enough to be used for the Selection approach.

\subsubsection{Number of corresponding colour patches for colour matching (NoCCP-CM)}

\begin{figure}[htbp]
    \centering
    \includegraphics[width=1.0\linewidth]{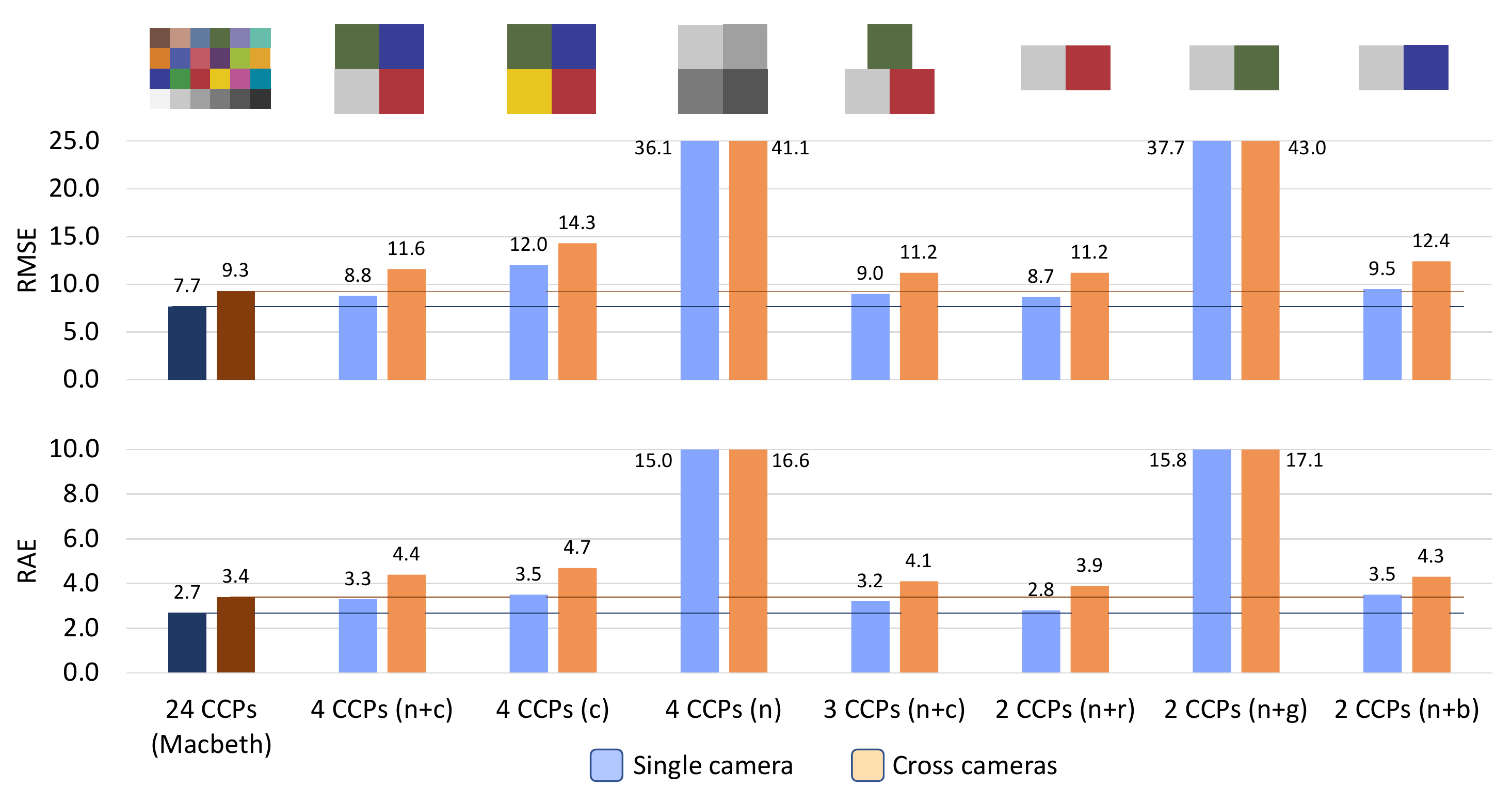}
    \caption{Single and cross camera RCC performance in terms of the RMSE and RAE using different number and composition of CCPs for colour matching. In this figure, $n$ stands for neutral, $c$ represents colourful such as red ($r$), green ($g$), and blue ($b$) CCPs. Colour matching using all the colour patches on a Macbeth chart (24 CCPs) was the comparison baseline. The CCPs used for each matching are indicated on top of the diagram where each coloured square represents an applied CCP.}
    \label{fig:nocp-cm}
\end{figure}

In this experiment, the effect and performance of applying different NoCCP-CM and the selection of which colour patches to use were examined. Experimental results were produced from the BoLD-alignment while applying different number of and composition of CCPs for colour matching yet keeping NoCI fixed at eight and NoCCP-CA fixed at 24. Applying the colour matching step that uses all the colour patches on a Macbeth chart as CCPs was assumed to provide an upper limit performance, i.e., the best possible, and was the comparison baseline. As shown in Fig. \ref{fig:nocp-cm}, the results generated by using three, four, and 24 Macbeth CCPs indicated that when a larger NoCCP-CM was used it generally led to higher RCC performance. As CCPs are expensive to be obtained, a balance between the NoCCP-CMs and the RCC performance needs to be struck. Based on the experimental results, a two mixed-colour CCPs (n=2) is the best option when CCPs for colour matching need to be limited as it still gives a satisfactory performance, even though it strongly depends on the selection of CCP colour composition. 24 Macbeth CCPs (n$\ge$4) is the better option when the performance and stability of the algorithm take priority.

\subsubsection{Selection of colour patches}

\begin{equation}\label{equ:br-chromaticity-ratio}
    ratio_{rb} =
    \begin{cases}
        \frac{b}{r}, b>r \\
        \frac{r}{b}, r>b
    \end{cases}
    \in \left[ {1,+\infty} \right)
\end{equation}

The colour matching performance strongly depends on the careful selection of the CCPs. When looking at two mixed-colour CCPs that used red, green, and blue as the colourful CCP, the performance generated with a red or a blue CCP were satisfactory yet unacceptable for that generated with a green CCP. This is due to the large BR chromaticity ratios (Eq. \ref{equ:br-chromaticity-ratio}) for the red (2.90) and blue (2.67) CCPs and small ratio in the case of a green (1.27) CCP. This indicated that CCPs with larger BR chromaticity ratio led to better correction accuracy. Comparison of different compositions of CCPs (neutral or colourful patches only and mixed patches) in Fig. \ref{fig:nocp-cm} illustrated that a mix of neutral and colourful CCPs improved the RCC performance. This is due to the larger colour intensity and chromaticity covering range they provide. These findings w.r.t CCP selection also apply to CCPs for camera calibration.

\subsection{Ablation studies}

\begin{figure}[htbp]
    \centering
    \includegraphics[width=\linewidth]{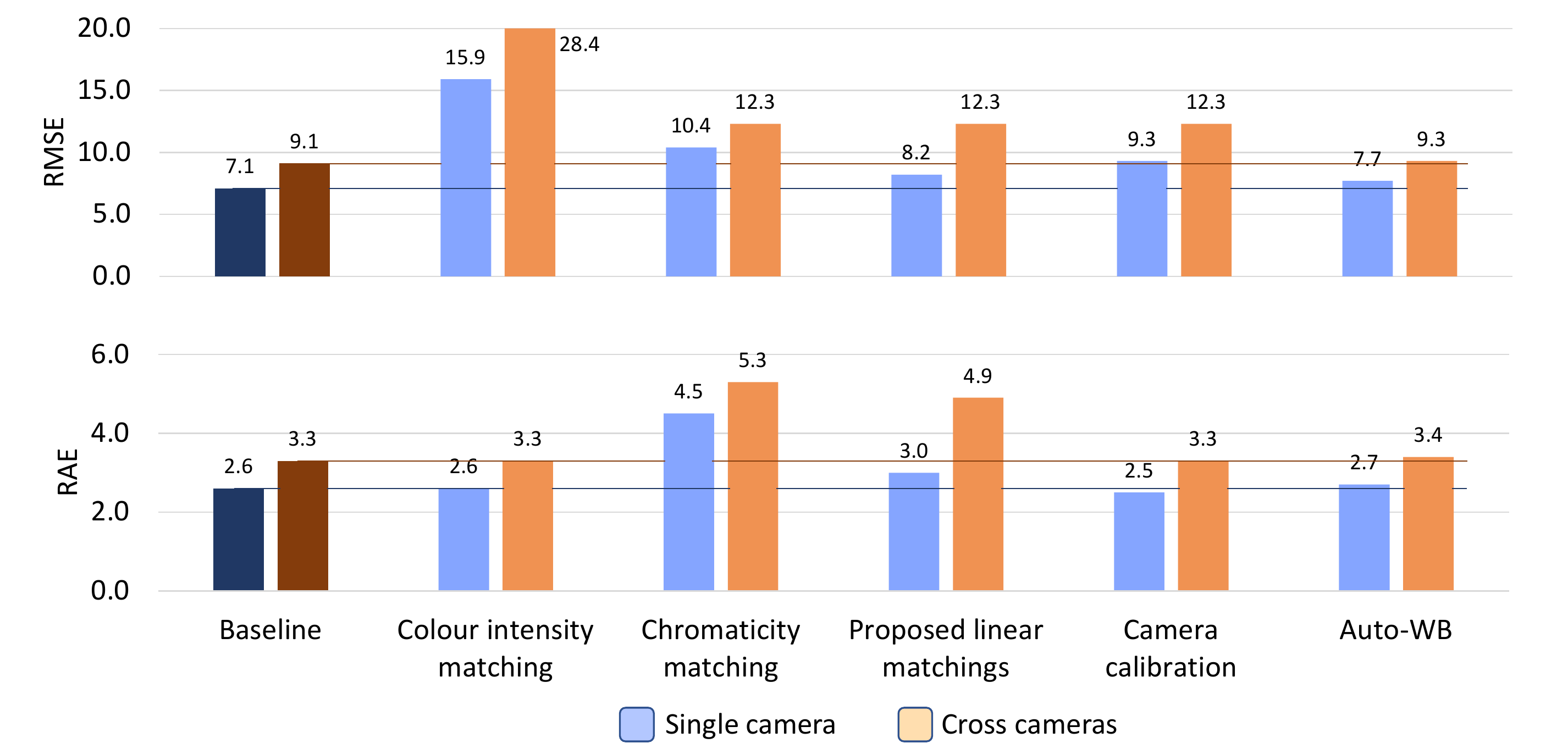}
    \caption{Single (blue bars) and cross (orange bars, i.e., colour alignment between images taken by different cameras) camera RCC performance contributions in terms of the RMSE and RAE of the different components in the BoLD-alignment are evaluated, compared and discussed in the text.}
    \label{fig:ablation}
\end{figure}

In the ablation study, the effectiveness of the four major components of the proposed BoLD-alignment, i.e., camera calibration, colour intensity matching, chromaticity matching, the new linear matchings, were studied. The effectiveness of each component was evaluated by removing that component to visualise each component’s contribution to the overall performance by comparing the resulting performance with that of the baseline. The Belfast dataset was used for conducting these studies. The baseline performance was evaluated with all processing steps applied except for a WB. 

\subsubsection{Ablation study on the colour intensity matching}

The test disabled the colour intensity matching and resulted in a large increase in RMSE. It indicated that colour intensity matching contributed greatly to the accuracy of colour intensity correction ($+3.93$ RMSE). 

\subsubsection{Ablation study on the chromaticity matching}

This test, with chromaticity matching disabled, resulted in a dramatic increase in RAE. This validated the importance of chromaticity matching on chromaticity correction accuracy ($+3.93$ RAE). 

\subsubsection{Ablation study on the proposed linear matchings}

In this test, conventional linear matchings were performed without utilising our proposed linear matching algorithms. The performance reduction compared to baseline highlights the important contribution of our proposed algorithms ($+4.31$ RMSE and $+2.02$ RAE). 

\subsubsection{Ablation study on the camera response calibration}

This test disabled the camera response calibration and that resulted in an increased RMSE showing the calibration’s contribution to colour intensity correction accuracy. By including the camera response calibration, the colour intensity correction performance improved but had little effect on that of chromaticity ($+5.32$ RMSE and $-0.03$ RAE). This is because the proposed model only corrects nonlinearity in colour intensity.

\subsubsection{Ablation study on the WB}

The final test enabled the WB step prior to applying the proposed BoLD-alignment. Its results did not show significant difference in accuracies ($+0.75$ RMSE and $+0.15$ RAE). This indicated that a WB ahead of the proposed model had little influence on the model performance.

\subsection{Benchmarks}

Two benchmarks were conducted to compare RCC performance in terms of the RMSE, RAE, $\Delta$E 2000, and execution times achieved by popular, state-of-the-art, and our proposed methods. 

\subsubsection{Single and cross camera RCC performance benchmark in terms of the RMSE, RAE, and $\Delta$E 2000}

\begin{table*}[htbp]
    \centering
    \caption{Single and cross camera RCC performance benchmark evaluated in terms of the RMSE (in intensity), RAE (in degree), and $\Delta$E 2000 using the three image datasets. Our BoLD-alignment achieved the best performance.}
    \scalebox{0.7}{
        \begin{tabular}{m{2.1cm}|m{1.9cm}!{\vrule width1pt}m{0.7cm}|m{0.7cm}|m{0.7cm}|m{0.7cm}|m{0.7cm}|m{0.7cm}!{\vrule width1pt}m{0.7cm}|m{0.7cm}|m{0.7cm}|m{0.7cm}|m{0.7cm}|m{0.7cm}!{\vrule width1pt}m{0.7cm}|m{0.7cm}|m{0.7cm}|m{0.7cm}|m{0.7cm}|m{0.7cm}}
            \toprule[1pt]
            \multicolumn{2}{c!{\vrule width1pt}}{\Large Method}    &  
            \multicolumn{6}{c!{\vrule width1pt}}{\large \makecell[c]{Created Belfast\\(Common scenes)}}       & 
            \multicolumn{6}{c!{\vrule width1pt}}{\large \makecell[c]{Modified Middlebury\\(Uncommon scenes)}}   & 
            \multicolumn{6}{c}{\large \makecell[c]{Rendered Gehler-Shi\\(Common scenes)}}    \\
            \hline
            \multirow{2}{*}{Name}  &  \multirow{2}{*}{Parameters}  &  \multicolumn{2}{c|}{RMSE}  & \multicolumn{2}{c|}{RAE} & \multicolumn{2}{c!{\vrule width1pt}}{$\Delta$E 2000} &  \multicolumn{2}{c|}{RMSE}  & \multicolumn{2}{c|}{RAE} & \multicolumn{2}{c!{\vrule width1pt}}{$\Delta$E 2000}     &  \multicolumn{2}{c|}{RMSE}  & \multicolumn{2}{c|}{RAE} & \multicolumn{2}{c}{$\Delta$E 2000} \\
            \cline{3-20}
             & & Single          & Cross        &  Single     & Cross   & Single     & Cross        & Single     & Cross   & Single     & Cross  & Single     & Cross           & Single     & Cross   & Single     & Cross  & Single   & Cross \\
            \midrule[1pt]
            \multicolumn{20}{c}{\large Baseline}                  \\
            \hline
            \multicolumn{2}{c!{\vrule width1pt}}{Original}         & 18.36  &  21.75  &  4.42  &  5.24  &  7.58  &  8.45         &  30.40 & 32.89 & 5.60 & 7.95 & 10.63 & 11.98           &  39.76 &  43.12  &  7.21  &  9.01  &  14.46  &  16.32  \\
            \midrule[1pt]
            \multicolumn{20}{c}{\large White Balance}               \\
            \hline
            \multicolumn{2}{c!{\vrule width1pt}}{White-Patch \cite{Ebner2007}}      & 17.99  &    21.57  &  4.40  & 5.19   & 7.30  &  8.33       & 25.41   & 27.86  & 6.86  & 8.55 & 8.24 & 9.33         &  36.45  &  40.18   &  6.93  &  8.83  &  13.29   &  15.16  \\
            \hline
            \multicolumn{2}{c!{\vrule width1pt}}{Grey-World \cite{Ebner2007}}        & 23.67  &    26.38  &  6.31  &  6.80  & 9.86  &  10.66       & 34.89  & 33.44  & 14.67 & 14.32  & 10.35  & 10.05         &  38.92  &  40.18  &  8.61  & 9.64   &  14.45  &   14.86  \\
            \hline
            \multicolumn{2}{c!{\vrule width1pt}}{Shades-of-Grey \cite{Finlayson2004}}  & 23.36   &   26.07  & 6.15  &  6.66  &  9.95  &  10.69           & 42.70  &  41.58  & 8.49  &  9.23  & 13.47  & 13.00           & 61.82   &  60.47  & 8.16  &  9.01   &  19.58   &  19.08    \\
            \hline
            \multicolumn{2}{c!{\vrule width1pt}}{Double-Opponency \cite{Gao2015}}      & 20.26   &   23.93  &  5.88  &  6.78   & 8.61    &   9.37           &  40.02  &  39.35  & 6.38 & 8.74  & 14.29  &  14.48          &  46.19   &  50.04  &  8.92   &  10.93  &  17.93   & 19.63    \\
            \hline
            \multicolumn{2}{c!{\vrule width1pt}}{CNN \cite{Bianco2015}} &  24.87   &  28.90    &  7.84  &  8.54   &  8.91   &  9.98                  & 32.12  &  31.44  &  10.61  &  12.02   &  11.12  &  10.73           &  45.00  &  48.63  &  11.14  &  13.25  &  15.97  & 17.76 \\
            \hline
            \multicolumn{2}{c!{\vrule width1pt}}{FC4 \cite{Hu2017}}  &  16.92 & 20.79   &  4.11  & 4.92   &  7.08  & 8.09             &  27.83  &  28.37   & 4.70  &  6.87   &  10.01  &  10.18          & 30.68  &  32.96    &  5.38   &  6.57  &  11.08  &  12.14  \\
            \hline
            \multicolumn{2}{c!{\vrule width1pt}}{sRGB \cite{Afifi2019}}   & 20.13  &  23.84   & 5.27 & 6.10   & 7.78 &  8.99          & 23.98  & 26.29   & 4.71  &  6.36  &  8.23  & 9.16         & 31.92  &  32.12   & 5.14  &  5.41  &  10.90  &  10.90   \\
            \hline
            \multicolumn{2}{c!{\vrule width1pt}}{Deep-Editing \cite{Afifi2020}}   & 16.05 & 19.45  & 4.21 & 4.55 & 6.51 & 7.59          & 27.67  & 26.39  &  7.00  & 7.08  & 11.12  &  10.67           & 32.76   &  33.06   &  4.81  &  5.31  &  11.46  &  11.80  \\
            \hline
            \multicolumn{2}{c!{\vrule width1pt}}{C5 \cite{Afifi2021} (based on CCC \cite{Barron})}   & 14.96  & 19.25  & \textbf{2.60} & \textbf{3.45}  & 5.53 & 6.71           & 23.82  &  27.79  &  3.59  &  5.15  & 8.70  &  9.89         & 27.96   &  30.63  &  \textbf{4.20}  &  \textbf{4.88}  & 9.67  &  10.56    \\
            \midrule[1pt]
            \multicolumn{20}{c}{\large Colour matching}       \\
            \hline
            Poly. \cite{Cheung2004} & \scriptsize \makecell[l]{ Terms: 3/5, \\ NoCCP-CM: 3 }    & 15.66     &  21.33  &  3.58  & 4.94  &  5.28  &  6.68                  & 23.66     &  23.17   & 5.29  & 6.71  &  6.88  &  6.90            &   29.60  &  46.14   &  10.66  &  9.12   &  14.48   &   16.29   \\
            \hline
            Root-Poly. \cite{Finlayson2015} & \scriptsize \makecell[l]{ Degree: 1, \\ NoCCP-CM: 3 }   & 15.66  &   21.33  &  3.58  &  4.94  &  5.28  &  6.68           & 23.66  &   23.17  &  5.29  &  6.71  &  6.88  &  6.90           &  32.04  & 89.94  &  5.56  &  20.69  & 9.18  &  21.41   \\
            \hline
            Vander. \cite{Westland2012} & \scriptsize \makecell[l]{ Degree: 1, \\ NoCCP-CM: 3  }   & 15.64    &   21.28  &  3.58  &  4.93  & 5.28  & 6.68                   & 23.66    &   23.17  &  5.29  &  6.71   & 6.88  &  6.90           &  32.00  &  89.88   &  5.56  &  20.62   & 9.18  &  21.40   \\
            \midrule[1pt]
            \multicolumn{20}{c}{\large Proposed}      \\
            \hline
            Our BoLD  & \scriptsize \makecell[l]{ NoCI: 4, \\ NoCCP-CA: 24, \\ NoCCP-CM: 2 }   &  \textbf{10.74} &  \textbf{15.41} & 2.77 &  3.91 &  \textbf{4.01} &  \textbf{5.23}          & \textbf{12.38} & \textbf{17.52} & \textbf{2.41} & \textbf{4.58}  & \textbf{3.89}   & \textbf{5.59}         & 21.42  &  22.61   &  4.78  &  5.90   &   \textbf{7.25}  &  \textbf{7.96}  \\
            \hline
            Our BoLD & \scriptsize \makecell[l]{ NoCI: 4, \\ NoCCP-CA: 24, \\ NoCCP-CM: 3 }     &  11.31  &  15.85 & 3.20 & 4.30   &  5.00  &   6.21          & 12.72 & 17.41 & 2.67 & 4.90  &  4.43  & 6.38         &  \textbf{20.57}  & \textbf{22.01}   &  4.80  &  6.09   & 7.64  &  8.83   \\
            \bottomrule[1pt]
        \end{tabular}
    }
    \label{tab:RCC-benchmark}
\end{table*}

Table \ref{tab:RCC-benchmark} presents the results of the benchmark where the proposed BoLD-alignment was compared with the other popular and state-of-the-art methods. The best performance of each metric is highlighted in bold. 

The first item listed in Table \ref{tab:RCC-benchmark} is the evaluated RCC performance on the original image datasets without applying any correction. This acted as the baseline for the comparisons.

Subsequently, eight WB methods that correct images to a neutral illumination were compared in the White Balance category of the benchmark. Four neural network-related methods were included in this category. Default gamma correction ($\gamma=2.2$) was applied in each WB method to try to transfer images from sRGB to RAW. However, Most results in this category did not demonstrate significant better performance than baseline. An explanation is the failure of scene assumptions used in these methods, e.g., the image is assumed grey in average by the Grey-World yet the images in the Belfast dataset, for example, have an averaged 4.29 RAE from the closest neural colours which could lead to a miscalculated illumination colour that then distorted the correction. Another explanation is that most of the WB methods are designed to work on RAW images rather than on the sRGB images despite the default gamma correction applied in this benchmark. And the colour distortions due to image post-processing procedures in imaging pipelines are complex and vary from camera to camera. This fact makes the images hard to be corrected by simple WB. C5 \cite{Afifi2021} produced very well RCC performance in terms of RAE, especially on the two image datasets with common scenes, by constructing a filter bank and $uv$ histograms during analysis.

In the Colour Matching category, the methods by Cheung et al. \cite{Cheung2004} and Finlayson et al. \cite{Finlayson2015} and the Vandermond method \cite{Westland2012} used three CCPs (Colour patch number 1, 9 and 11 on the Macbeth chart). The explanation for their results is the inadequate nonlinear mapping that results from the limit in NoCCP-CMs (n=3).

By comparison, our method that used only two (Colour patch number 1 and 9) or three (Colour patch number 1, 9 and 11 on the Macbeth chart) NoCCP-CMs achieved the best overall RCC performance on the three datasets, especially on the modified Middlebury dataset of uncommon scenes that imitates scientific imaging scenes. It outperformed the tested statistical and learning-based WB methods that considered only the illuminations. It also achieved superior performance than colour matching methods that used identical or even larger NoCCP-CMs.

\subsubsection{Computation time benchmark for calibration and correction}

\begin{table}[htbp]
    \centering
    \scalebox{0.8}{
        \begin{tabular}{c|c}
            \toprule[1pt]
            \large Method                   & \large \makecell[c]{ Execution time \\in seconds } \\
            \midrule[1pt]
            \multicolumn{2}{c}{\large Response calibration} \\
            \hline
            Reference                 & 0.005                     \\
            Exposure                  & 165.77                    \\
            Our Selection approach    & 3.50                      \\
            Our Optimisation approach & 1812.33                   \\
            \midrule[1pt]
            \multicolumn{2}{c}{\large Colour correction} \\
            \hline
            White-Patch                 & 0.22                     \\
            Grey-World                  & 0.31                    \\
            Shades-of-Grey           & 0.62                      \\
            Double-Opponency        & 0.31                   \\
            \hline
            CNN           &  9.18                      \\
            FC4        &    7.06                   \\
            sRGB        &    2.03                   \\
            Deep-Editing     &    5.80                \\
            C5        &    1.4                   \\
            \hline
            Ours        &    0.92                   \\
            \bottomrule[1pt]
        \end{tabular}
    }
    \vspace{5pt}
    \caption{Execution time (in seconds) benchmark of the popular and proposed calibration (i.e., camera response calibration) and colour correction (i.e., response linearisation and colour matching) methods.}
    \label{tab:calibration-execution-time}
\end{table}

This benchmark compared the time taken to calibrate each camera and correct an image. Table \ref{tab:calibration-execution-time} presents the results of this benchmark. 

The Reference and Exposure camera response calibration approaches are used for comparison purposes. In the Reference approach that polynomial regressed (d=6) on the standard reference values accomplished within 0.005s. It was extremely fast due to the well-optimised and relatively simple polynomial regression code implementation. The Exposure calibration approach calculated the optimal iCRF From a pre-set exposure times and machine learning based optimisation. In contrast to the relatively fast Reference approach, it took 165.77s for the Exposure approach to complete. For our proposed camera response calibration, the Selection approach took only 3.50s to accomplish the task compared to 1812.33s taken by the Optimisation approach.

In the second section it shows that our method (0.92s) performed slightly slower than the statistical-based WB methods due to the nonlinear interpolation during the response linearisation, yet faster than the learning-based methods.

\section{Discussions}

When it comes to the choosing between the Selection and Optimisation calibration approaches, the experimental results have unequivocally indicated that the Selection approach almost always outperformed the Optimisation approach in its accuracy, precision, and execution speed. Hence, the Selection approach is almost certainly to be the better option.

Even though colour charts were imaged and utilised for the ease of model validation and evaluation, the proposed colour alignment model can certainly work with non-standard colour references that are more accessible and easier-to-operate than the colour charts. An automatic CCP detection method can also be added to the model to further improve its applicability.

Finally, it can be said that the tests and benchmarks demonstrate that the BoLD feature works as a procedure for camera response calibration. Given its generality there is much potential for applications in other fields where nonlinearity needs to be measured and calibrated without the availability of standard references.

\section{Conclusion}

In this paper, a high-performance colour alignment model that aligns colour of images taken by different cameras and under varied illuminations has been proposed. The model consists of three steps: camera response calibration, response linearisation, and colour matching. The initial step needs only be performed once for each device and is applicable to commercial digital cameras that do not give access to internal imaging sensor data. Access to standard colour references is not required. The second step works much faster than a neural network based method and is suitable to be used even for portable devices, e.g., smartphones. While in the third step, only two CCPs are needed as the minimum requirement to perform the colour matching. It improves its applicability as CCP is also expensive to obtain in consumer-oriented imaging applications. Overall, our proposed model has achieved the best RCC performance on two image datasets but still only requires an acceptable complexity in terms of execution time (3.50s for camera response calibration and 0.92s for correcting an image) in the benchmarks.

\section{Acknowledgment}

This project was supported in part by the European Union’s Horizon 2020 research and innovation program under the Marie-Sklodowska-Curie grant agreement No 720325, FoodSmartphone, and in part by the Key Laboratory of Intelligent Preventive Medicine of Zhejiang Province 2020E10004.

\newcommand\BIBentryALTinterwordstretchfactor{
    2.5}
\bibliographystyle{IEEEtran}
\bibliography{IEEEabrv,References/library.bib}

\end{document}